\documentclass[10pt]{iopart}
\usepackage{graphicx}
\usepackage{iopams}

\newcommand{\sgn}{\mathop{\mathrm{sgn}}\nolimits}

\newcommand{\vect}[1]{\bi{#1}}
\newcommand{\journaltitle}[1]{\textit{#1}}

\newcommand{\volume}[1]{\textbf{#1}}

\newcommand{\Real}{\mathop{\mathrm{Re}}\nolimits}

\newcommand{\half}{{\textstyle \frac{1}{2}}}
\newcommand{\threehalf}{{\textstyle \frac{3}{2}}}
\newcommand{\thalf}{{\scriptstyle \frac{1}{2}}}

\newcommand{\tfourth}{{\scriptstyle \frac{1}{4}}}
\newcommand{\tthreequarter}{{\scriptstyle \frac{3}{4}}}

\def\hbarit {{\mathchar'26\mkern-11muh}} 
\begin{document}
\jl{1}
\title{Time correlations in a confined magnetized free-electron gas}
\author{L G Suttorp}
\address{Instituut voor Theoretische Fysica, Universiteit van Amsterdam, 
Valckenierstraat 65, 1018 XE Amsterdam, The Netherlands}

\begin{abstract}
The time-dependent pair correlation functions for a degenerate ideal quantum gas
of charged particles in a uniform magnetic field are studied on the basis of
equilibrium statistics. In particular, the influence of a flat hard wall on the
correlations is investigated, both for a perpendicular and a parallel orientation
of the wall with respect to the field. The coherent and incoherent parts of the
time-dependent structure function in position space are determined from an
expansion in terms of the eigenfunctions of the one-particle Hamiltonian. For
the bulk of the system, the intermediate scattering function and the dynamical
structure factor are derived by taking successive Fourier transforms. In the
vicinity of the wall the time-dependent coherent structure function is found to
decay faster than in the bulk. For coinciding positions near the wall the form
of the structure function turns out to be independent of the orientation of the
wall. Numerical results are shown to corroborate these findings.
\end{abstract}   


\submitted 

\maketitle

\section{Introduction}

The equilibrium properties of magnetized quantum electron gases are partly
determined by physical processes near its boundaries. For instance, the Landau
diamagnetic effect of a magnetized free-electron gas is a result of currents
flowing near the edge \cite{BOH:1911}. Hence, it is important to study the
influence of a wall on both the static and the dynamic behavior of these
systems. In a recent paper \cite{KES:2001} we investigated the static
correlations in a fully degenerate magnetized free-electron gas in the vicinity
of a wall. In the bulk the decay of the pair correlation function for large
separations depends on the orientation of the position difference: for
directions orthogonal to the field it is exponentially fast, whereas it is
algebraic for directions parallel to the field. If a hard wall parallel to the
field is present, the picture changes: the asymptotic behavior of the pair
correlation function for transverse directions changes from exponential to
algebraic for any two positions close to the wall.

Further insight in the influence of a wall on the correlation properties is
obtained by studying dynamical correlations, which are the subject of the
present paper. In the bulk these time correlations yield, after suitable Fourier
transforms, the dynamic structure factor, which has been studied quite
extensively for various systems, both theoretically and experimentally. For an
unmagnetized degenerate free-electron gas its properties were determined several
decades ago \cite{LIN:1954}. The generalization to finite temperatures is
available as well \cite{KHA:1976}. The dynamic structure factor for a magnetized
degenerate free-electron gas is somewhat more complicated, as it involves a sum
over Landau levels. If a wall is present as well, the translation invariance of
the system is lost, so that taking a Fourier transform is inconvenient. Instead,
it is more suitable to analyze the dynamical correlations by means of the
structure function in configuration space. We will focus on its time dependence,
and, in particular, on its decay for large time. As a starting point we shall
employ the expansion of the space-time-dependent structure function in terms of
the single-particle eigenfunctions.

The paper is organized as follows. In the next section we shall define the
structure function and the associated time correlation functions. In section 3
we shall derive expressions for the bulk space-time-dependent structure function
of a magnetized degenerate free-electron gas. Some numerical results for this
function will be presented as well. The sections 4 and 5 contain an analysis of
the intermediate scattering functions and the dynamical structure factor for the
bulk, again with some numerical illustrations. In section 6 we shall verify that
in the limit of vanishing magnetic field our results reduce to the well-known
expressions for the field-free degenerate free-electron gas. Edge effects are
the subject of sections 7 and 8. Both the cases of a wall perpendicular to and
parallel with the magnetic field will be discussed. The analytical results will
be checked numerically.

\section{General properties of quantum time correlation functions}

The equilibrium time-dependent structure function in position space is defined
as
\begin{equation}
\fl S({\bf r},{\bf r}',t)=\left\langle \rme^{\rmi Ht}\psi^{\dagger}({\bf r})
\psi({\bf r})\rme^{-\rmi Ht}\; \psi^{\dagger}({\bf r}')
\psi({\bf r}')\right\rangle 
-\langle \psi^{\dagger}({\bf r}) \psi({\bf r})\rangle\; 
\langle \psi^{\dagger}({\bf r}') \psi({\bf r}')\rangle\; 
\label{2.1}
\end{equation}
with $\psi({\bf r})$ the field annihilation operator and $H$ the many-body
Hamiltonian. The brackets denote an equilibrium average in the grand-canonical
ensemble, with the inverse temperature $\beta$ and the chemical potential
$\mu$. We have chosen units such that $\hbarit$ drops out. 

For independent particles with Fermi statistics the structure function can
be expressed in terms of single-particle wavefunctions as
\begin{eqnarray}
S({\bf r},{\bf r}',t)=\sum_j \sum_{j'}\frac{1}{\rme^{\beta(E_j-\mu)}+1}\;
\left[1-\frac{1}{\rme^{\beta(E_{j'}-\mu)}+1}\right]\nonumber\\
\times\varphi^{\ast}_j({\bf r})\varphi_j({\bf r}')\varphi_{j'}({\bf r})
\varphi^{\ast}_{j'}({\bf r}')\; \rme^{\rmi(E_j-E_{j'})t}
\label{2.2}
\end{eqnarray}
where $\varphi_j({\bf r})$ are the eigenfunctions of the single-particle
Hamiltonian with eigenvalues $E_j$. The structure function can be written as the
sum of an incoherent and a coherent part \cite{HOV:1954}. These are given as
\begin{eqnarray}
S_{inc}({\bf r},{\bf r}',t)=G_{\mu,T}^{\ast}({\bf r},{\bf r}',t)\; 
G({\bf r},{\bf r}',t)
\label{2.3}\\
S_{coh}({\bf r},{\bf r}',t)=-\left|G_{\mu,T}({\bf r},{\bf r}',t)\right|^2
\label{2.4}
\end{eqnarray}
Here we introduced the time correlation functions $G_{\mu,T}$ and $G$, which are
defined as
\begin{eqnarray}
G_{\mu,T}({\bf r},{\bf r}',t)=\sum_j \frac{1}{\rme^{\beta(E_j-\mu)}+1}\; 
\varphi_j({\bf r})\varphi^{\ast}_j({\bf r}')\; \rme^{-\rmi E_j t}
\label{2.5}\\
G({\bf r},{\bf r}',t)=\sum_j \varphi_j({\bf r})\varphi^{\ast}_j({\bf r}')
\; \rme^{-\rmi E_j t}
\label{2.6}
\end{eqnarray}
The first function depends on $\mu$ and $T$, whereas the second does not. In the
static case of vanishing $t$ the correlation function $G$ reduces to
$\delta({\bf r}-{\bf r}')$.

Our main interest in the following will be in the fully degenerate case with
$T=0$. In that case $G_{\mu,T}$ reduces to
\begin{equation}
G_{\mu,T=0}({\bf r},{\bf r}',t)=\sum_j \theta\left(\mu-E_j\right)\; 
\varphi_j({\bf r})\varphi^{\ast}_j({\bf r}')\; \rme^{-\rmi E_j t}
\label{2.7}
\end{equation}
with $\theta$ the step function. The time correlation function for finite
$T$ can easily be recovered from that of the fully degenerate case, since
once has 
\begin{equation}
G_{\mu,T}({\bf r},{\bf r}',t)=\int_0^{\infty}d\mu'\; 
\frac{1}{\rme^{\beta(\mu'-\mu)}+1}\;
\frac{d}{d\mu'}\; G_{\mu',T=0}({\bf r},{\bf r}',t)
\label{2.8}
\end{equation}
The subscript $T=0$ will often be omitted in the following, so that
$G_{\mu,T=0}$ will be written as $G_{\mu}$.

In uniform systems, or in the bulk of a confined system, the structure function
$S({\bf r},{\bf r}',t)$ depends on the position difference ${\bf r}-{\bf r}'$
only. In this case the intermediate scattering function follows by taking the
spatial Fourier transform of $S$:
\begin{equation}
S({\bf k},t)=\int d({\bf r}-{\bf r}')\; 
\rme^{-\rmi{\bf k}\cdot({\bf r}-{\bf r}')}\; S({\bf r},{\bf r}',t)
\label{2.9}
\end{equation}
and similarly for the incoherent and coherent parts. A further Fourier transform
leads to the dynamic structure factor:
\begin{equation}
S({\bf k},\omega)=\int dt\; \rme^{\rmi\omega t}\; S({\bf k},t)
\label{2.10}
\end{equation}
Substituting (\ref{2.5}) and (\ref{2.6}) in (\ref{2.3}) and performing the
Fourier transforms one gets
\begin{eqnarray}
\fl S_{inc}({\bf k},\omega)=2\pi\sum_j \sum_{j'}
\frac{1}{\rme^{\beta(E_j-\mu)}+1}\; 
\delta\left(\omega+E_j-E_{j'}\right)\nonumber\\
\times\int d({\bf r}-{\bf r}')\; 
\rme^{-\rmi{\bf k}\cdot({\bf r}-{\bf r}')}\; 
\varphi^{\ast}_j({\bf r})\varphi_j({\bf r}')\varphi_{j'}({\bf r})
\varphi^{\ast}_{j'}({\bf r}')
\label{2.11}
\end{eqnarray}
A similar expression can be written down for the coherent part. Comparing the
two expressions one easily finds the relation
\begin{equation}
S_{coh}({\bf k},\omega)=
\frac{1}{\rme^{\beta\omega}-1}\; S_{inc}({\bf k},\omega)
+\frac{1}{\rme^{-\beta\omega}-1}\; S_{inc}(-{\bf k},-\omega)
\label{2.12}
\end{equation}
This relation also follows from the principle of detailed balance
\cite{PIN:1966}, if the symmetry of the coherent dynamic structure factor under
a change of sign of both ${\bf k}$ and $\omega$ is taken into account.

\section{Time correlations in the bulk}

We consider a system of charged particles which move in a uniform magnetic field
directed along the $z$-axis. The interaction between the particles is
neglected. To describe the magnetic field we adopt the Landau gauge, with vector
potential $\vect{A}=(0,Bx,0)$. The single-particle Hamiltonian reads
\begin{equation}
H=-\half\Delta +\rmi Bx\frac{\partial}{\partial y} +\half B^2 x^2
\label{3.1}
\end{equation}
Units have been chosen such that the charge and the mass of the particles drop 
out, while $c$ (and $\hbarit$) have been put to 1 as well. From now on we will 
often measure distances in terms of the cyclotron radius $1/\sqrt{B}$. To that 
end we introduce the reduced variables $\xi=\sqrt{B}x$, $\eta=\sqrt{B}y$ and 
$\zeta=\sqrt{B}z$. 

The normalized eigenfunctions of the Hamiltonian are:
\begin{equation}
\fl \psi_{n_y,n_z,n}({\bf r})=\frac{B^{1/4}}{2^{n/2}\pi^{1/4}(n!)^{1/2}L} \;
H_n(\xi-\kappa_y)\; \rme^{\rmi \kappa_y \eta +\rmi \kappa_z\zeta-\thalf
(\xi-\kappa_y)^2}
\label{3.2}
\end{equation}
We imposed periodic boundary conditions in the $y$- and $z$-direction, with a
periodicity length $L$. The wavevector components are given by $k_i=2\pi n_i/L$,
with integer $n_i$ for $i=y,z$; the reduced wavevector components are
$\kappa_i=k_i/\sqrt{B}$.  The Hermite polynomials $H_n$ carry the non-negative
integer label $n$. The associated eigenvalue is the sum of the kinetic energy
for the $z$-direction and the Landau-level energy for the transverse directions:
\begin{equation}
E_{n_z,n}=B\left(n+\half\right)+\half k_z^2
\label{3.3}
\end{equation}

The time correlation function $G_{\mu}$ for the fully degenerate system follows by
inserting the eigenfunctions and eigenvalues in (\ref{2.7}). For large $L$ the
summations over $n_i$ may be replaced by integrations over $\kappa_i$. As a
result we find:
\begin{equation}
\fl G_{\mu}({\bf r},{\bf r}',t)=
\frac{B^{3/2}}{4\pi^{5/2}}\sum_{n=0}^{\infty}\frac{1}{2^nn!}
\rme^{-\rmi(n+\thalf)\tau}\; J_{\perp,n}(\xi,\xi',\eta,\eta')\; 
J_{\parallel,n,\nu}(\zeta,\zeta',\tau)
\label{3.4}
\end{equation}
We introduced the reduced time $\tau=Bt$, and the reduced chemical potential
$\nu=\mu/B$. The integrals over $\kappa_i$ are given by
\begin{eqnarray}
\fl J_{\perp,n}(\xi,\xi',\eta,\eta')=
\int_{-\infty}^{\infty}d\kappa_y\;
H_n(\xi-\kappa_y)\; H_n(\xi'-\kappa_y)\;\rme^{
\rmi\kappa_y(\eta-\eta')-\thalf(\xi-\kappa_y)^2-\thalf(\xi'-\kappa_y)^2}
\label{3.5}
\\
\fl J_{\parallel,n,\nu}(\zeta,\zeta',\tau)=
\int_{-\infty}^{\infty}d\kappa_z \;
\theta\left(\nu-\half\kappa_z^2-n-\half\right)\;
\rme^{\rmi\kappa_z(\zeta-\zeta')-\thalf\rmi\kappa_z^2\tau}
\label{3.6}
\end{eqnarray}
The integration over the transverse wavevector can be carried out by using the
identity \cite{ERD:1954}:
\begin{equation}
\int_{\infty}^{\infty}dk\; H_n(k+a)\; H_n(k+b)\; \rme^{-k^2}=2^n\sqrt{\pi} \; n!\;
L_n(-2ab)
\label{3.7}
\end{equation}
for arbitrary complex $a$ and $b$. One finds:
\begin{equation}
\fl J_{\perp,n}(\xi,\xi',\eta,\eta')=2^n\sqrt{\pi}\; n!\; 
\rme^{-\tfourth
(\brho_{\perp}-\brho_{\perp}')^2
+\thalf\rmi(\xi+\xi')(\eta-\eta')}\;
L_n\left[\half(\brho_{\perp}-\brho_{\perp}')^2\right]
\label{3.8}
\end{equation}
with the two-dimensional vectors $\brho_{\perp}=(\xi,\eta)$ and
$\brho_{\perp}'=(\xi',\eta')$
The integral over the longitudinal wavevector can be rewritten in terms of
Fresnel integrals \cite{GRAD:1980}:
\begin{equation} 
\fl
J_{\parallel,n,\nu}(\zeta,\zeta',\tau)=\theta\left(\nu-n-\half\right)
\sqrt{\frac{\pi}{\tau}}\; \rme^{\thalf\rmi(\zeta-\zeta')^2/\tau}\; 
F\left[\sqrt{\tau(\nu-n-\thalf)},\frac{\zeta-\zeta'}{\sqrt{2\tau}}\right]
\label{3.9}
\end{equation}
with:
\begin{equation}
F(a,b)=C(a+b)-\rmi S(a+b)+C(a-b)-\rmi S(a-b)
\label{3.10}
\end{equation}
Combining (\ref{3.4}) with (\ref{3.8}) and (\ref{3.9}), we get
\begin{eqnarray}
\fl G_{\mu}({\bf r},{\bf r}',t)=
\frac{B^{3/2}}{4\pi^{3/2} \tau^{1/2}} \;
\rme^{-\tfourth(\brho_{\perp}-\brho_{\perp}')^2
+\thalf\rmi(\xi+\xi')(\eta-\eta')
+\thalf\rmi(\zeta-\zeta')^2/\tau}
 \nonumber\\ 
\times\; {\sum_n}' \rme^{-\rmi(n+\thalf)\tau}\;
 L_n\left[\half(\brho_{\perp}-\brho_{\perp}')^2\right]\; 
F\left[\sqrt{\tau(\nu-n-\thalf)},\frac{\zeta-\zeta'}{\sqrt{2\tau}}\right]
\label{3.11} 
\end{eqnarray}
The prime indicates that the summation is only over those values of $n$ for 
which $\nu-(n+\half)$ is non-negative, i.e. over the Landau levels that are at 
least partially filled. It should be noted that the right-hand side is
continuous as $\nu$ equals a half-odd integer, since $F(a,b)$ vanishes for
$a=0$. 

The function $G({\bf r},{\bf r}',t)$ can be evaluated along similar
lines. Instead of $J_{\parallel,n,\nu}(\zeta,\zeta',\tau)$ one encounters an
integral like (\ref{3.6}), without the step function, which is easily
evaluated. As a result we get
\begin{eqnarray}
\fl G({\bf r},{\bf r}',t)= \frac{(1-\rmi)B^{3/2}}{4\pi^{3/2} \tau^{1/2}} \;
\rme^{-\tfourth(\brho_{\perp}-\brho_{\perp}')^2
+\thalf\rmi(\xi+\xi')(\eta-\eta')
+\thalf\rmi(\zeta-\zeta')^2/\tau}
\nonumber\\ 
\times\; \sum_{n=0}^{\infty} \; \rme^{-\rmi(n+\thalf)\tau}\;
L_n\left[\half(\brho_{\perp}-\brho_{\perp}')^2\right]
\label{3.12}
\end{eqnarray}
Here the summation extends over all non-negative integer $n$. We should add a
small negative imaginary part to $\tau$ to ensure convergence. A simpler 
form for $G$ is obtained by using the identity \cite{ERD:1953}
\begin{equation}
\sum_{n=0}^{\infty}\; L_n(a)\; b^n=\frac{1}{1-b}\; \rme^{-ab/(1-b)}
\label{3.13}
\end{equation}
for $|b|<1$. In this way (\ref{3.12}) becomes
\begin{eqnarray}
\fl G({\bf r},{\bf r}',t)= -\frac{(1+\rmi)B^{3/2}}{8\pi^{3/2}
\tau^{1/2}\sin(\thalf\tau)} 
\rme^{\tfourth\rmi\cot(\thalf\tau)(\brho_{\perp}-\brho_{\perp}')^2
+\thalf\rmi(\xi+\xi')(\eta-\eta') +\thalf\rmi(\zeta-\zeta')^2/\tau}
\nonumber\\
\mbox{}
\label{3.14}
\end{eqnarray}
Clearly, $G$ is singular for $\tau$ equal to a multiple of $2\pi$.

Information on the equal-time correlations is obtained by putting $t=0$. The
function $G$ is equal to $\delta({\bf r}-{\bf r}')$ for $t=0$, as is obvious from its
definition. To determine the static limit of $G_{\mu}$ one uses the asymptotic
expansion \cite{GRAD:1980} for $u\rightarrow\infty$:
\begin{equation}
C(u)-\rmi\; S(u) \approx \half(1-\rmi)+\frac{\rmi}{\sqrt{2\pi}\; u}\; \rme^{-\rmi
u^2}+\ldots 
\label{3.15}
\end{equation}
In this way one finds from (\ref{3.11}):
\begin{eqnarray}
\fl G_{\mu}({\bf r},{\bf r}',t=0)=\frac{B^{3/2}}{2\pi^2|\zeta-\zeta'|}\;
\rme^{-\tfourth(\brho_{\perp}-\brho_{\perp}')^2
+\thalf\rmi(\xi+\xi')(\eta-\eta')}\nonumber\\
\times{\sum_n}'\; L_n\left[\half(\brho_{\perp}-\brho_{\perp}')^2\right]\;
\sin\left[\sqrt{2(\nu-n-\thalf)}|\zeta-\zeta'|\right]
\label{3.16}
\end{eqnarray}
This result agrees with that found before \cite{KES:2001}, if spin degeneracy is
taken into account.

The behavior of the time correlation functions depends on the orientation of
the position difference vector with respect to the magnetic field. In
particular, one may consider purely transverse and purely longitudinal
directions separately.

For a purely transverse position difference, e.g.\ for $\xi\neq\xi'$ with
$\eta=\eta'$ and $\zeta=\zeta'$, one finds from (\ref{3.11}):
\begin{eqnarray}
\fl G_{\mu}({\bf r},{\bf r}',t)=
\frac{B^{3/2}}{2\pi^{3/2} \tau^{1/2}} \;
\rme^{-\tfourth(\xi-\xi')^2}
\; {\sum_n}' \rme^{-\rmi(n+\thalf)\tau}\;
 L_n\left[\half(\xi-\xi')^2\right]\nonumber\\ 
\times\left[C\left(\sqrt{\tau(\nu-n-\thalf)}\right)
-\rmi S\left(\sqrt{\tau(\nu-n-\thalf)}\right)\right]
\label{3.17}
\end{eqnarray}
This expression gets particularly simple if only the lowest Landau level is
filled. For $\half\leq \nu \leq \threehalf$ one gets:
\begin{equation}
\fl G_{\mu}({\bf r},{\bf r}',t)=
\frac{B^{3/2}}{2\pi^{3/2} \tau^{1/2}} \;
\rme^{-\tfourth(\xi-\xi')^2-\thalf\rmi\tau}
\left[C\left(\sqrt{\tau(\nu-\half)}\right)-
\rmi S\left(\sqrt{\tau(\nu-\half)}\right)\right]
\label{3.18}
\end{equation}
The spatial decay is Gaussian, with a Laguerre-polynomial modulation in the
general case (\ref{3.17}). As a function of time, $G_{\mu}$ is algebraically
damped, with a periodic phase factor and a Fresnel-integral modulation. For
large $t$ the latter drops out, as follows from (\ref{3.15}). Owing to the
Laguerre polynomials in (\ref{3.17}), the asymptotic form of the coherent
structure function $S_{coh}({\bf r},{\bf r}',t)$ for large $t$ varies with the
magnitude of the transverse position difference, even if the Gaussian prefactor
is regarded as a trivial normalization factor. As we shall see, this feature
disappears when the position difference becomes purely longitudinal.

The behavior of $G$ for transverse separations is obtained from (\ref{3.14})
with $\eta=\eta'$ and $\zeta=\zeta'$ inserted:
\begin{equation}
G({\bf r},{\bf r}',t)= -\frac{(1+\rmi)B^{3/2}}{8\pi^{3/2}
\tau^{1/2}\sin(\thalf\tau)} \rme^{\tfourth\rmi\cot(\thalf\tau)(\xi-\xi')^2}
\label{3.19}
\end{equation}
The function depends on the position difference through a phase factor only. As
to the time dependence, the same singular behavior for $\tau= 2\pi m$ as in
(\ref{3.14}) is obtained.

For a purely longitudinal position difference, the behavior of $G_{\mu}$ is
found from (\ref{3.11}) by taking $\xi=\xi'$ and $\eta=\eta'$. One gets:
\begin{equation}
\fl G_{\mu}({\bf r},{\bf r}',t)=
\frac{B^{3/2}}{4\pi^{3/2} \tau^{1/2}} \;
\rme^{\thalf\rmi(\zeta-\zeta')^2/\tau}\;
 {\sum_n}' \rme^{-\rmi(n+\thalf)\tau}\;
F\left[\sqrt{\tau(\nu-n-\thalf)},\frac{\zeta-\zeta'}{\sqrt{2\tau}}\right]
\label{3.20} 
\end{equation}
The spatial dependence is no longer Gaussian: it is contained in a phase factor
and in the Fresnel integrals. As a function of time, a qualitatively similar
behavior as in the transverse case is found, with an algebraic damping, a
periodic phase factor and a Fresnel-integral modulation. For large $t$ the
latter disappears, as before, so that then $G_{\mu}$ depends on the position
difference through a phase factor only. Hence, for any value of the longitudinal
position difference, $S_{coh}({\bf r},{\bf r}',t)$ has the same asymptotic form
for large $t$.

The behavior of $G$ for longitudinal separations immediately follows from
(\ref{3.14}) as
\begin{equation}
G({\bf r},{\bf r}',t)= -\frac{(1+\rmi)B^{3/2}}{8\pi^{3/2}
\tau^{1/2}\sin(\thalf\tau)} 
\rme^{\thalf\rmi(\zeta-\zeta')^2/\tau}
\label{3.21}
\end{equation}
As in the transverse case, $G$ depends on the position difference through a
phase factor only. The time dependence is dominated by the usual
singularities.

Finally, let us consider the time dependence of $G_{\mu}$ and $G$ for coinciding
positions (${\bf r}={\bf r}'$). It is obtained by putting $\xi=\xi'$ in
(\ref{3.17}) and (\ref{3.19}):
\begin{eqnarray}
\fl G_{\mu}({\bf r},{\bf r},t)=
\frac{B^{3/2}}{2\pi^{3/2} \tau^{1/2}} \;
{\sum_n}' \rme^{-\rmi(n+\thalf)\tau}\;
\left[C\left(\sqrt{\tau(\nu-n-\thalf)}\right)
-\rmi S\left(\sqrt{\tau(\nu-n-\thalf)}\right)\right]\nonumber\\
\mbox{}\label{3.22}\\
G({\bf r},{\bf r},t)= -\frac{(1+\rmi)B^{3/2}}{8\pi^{3/2}
\tau^{1/2}\sin(\thalf\tau)} 
\label{3.23}
\end{eqnarray}
For large $t$, the function $G_{\mu}$ reduces to
\begin{equation}
\fl G_{\mu}({\bf r},{\bf r},t)\approx
\frac{(1-\rmi)B^{3/2}}{4\pi^{3/2} \tau^{1/2}} \;
{\sum_n}' \rme^{-\rmi(n+\thalf)\tau}
=\frac{(1-\rmi)B^{3/2}}{4\pi^{3/2}}\;
\frac{\sin\left[\thalf (n_0+1)\tau\right]}{\tau^{1/2}\sin(\thalf\tau)} \;
\rme^{-\thalf\rmi (n_0+1) \tau}
\label{3.24}
\end{equation}
The sine function in the numerator, with $n_0$ the label of the highest occupied
Landau level, ensures that $G_{\mu}$ is non-singular for $\tau=2\pi m$, in
contrast to $G$.

\begin{figure}
  \begin{center}
    \includegraphics[height=5.cm]{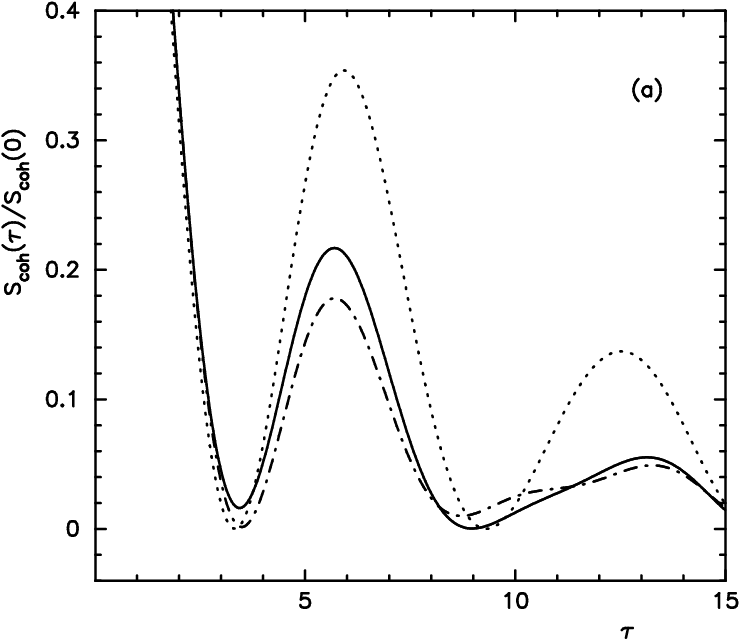}
    \hspace{0.5cm}
    \includegraphics[height=5.0cm]{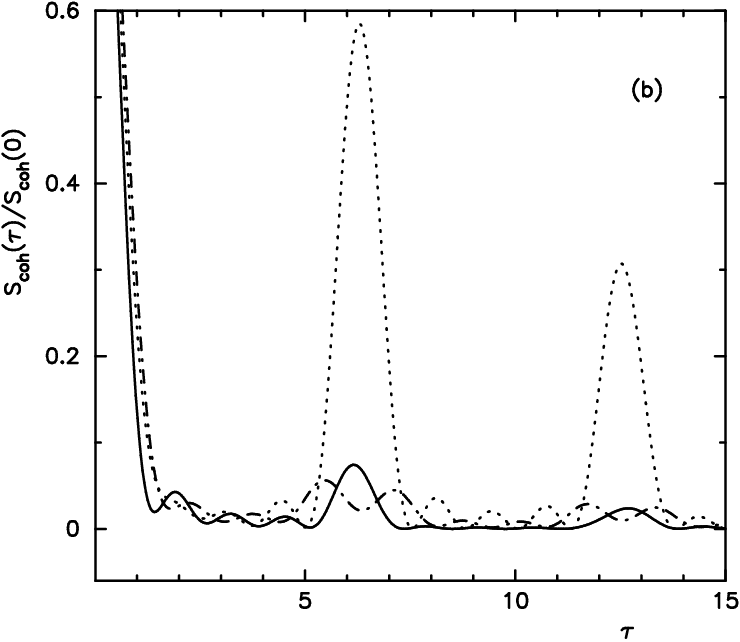}
  \caption{Bulk coherent structure function for $\nu=2$ (a) and $\nu=5$
    (b), as a function of time $\tau$. The curves, which are normalized to 1
    for $\tau=0$, represent the structure function for coinciding positions
    (\full), for a longitudinal position difference $|\zeta-\zeta'|=1$
    (\dotted), and for a transverse position difference $|\xi-\xi'|=1$
    (\chain).}
\label{fig1}
  \end{center}
\end{figure}

In figure \ref{fig1} we show numerical results for the normalized bulk coherent
structure function $S_{coh}({\bf r},{\bf r}',t)/S_{coh}({\bf r},{\bf r}',0)$, at
two values of the reduced chemical potential $\nu$. The anisotropy in the
behavior of the structure function for non-coinciding positions is clearly
visible. The curves that correspond to purely longitudinal position differences
show characteristic revivals around $\tau=2\pi m$, with integer $m$. The revivals are
less prominent for purely transverse position differences. These features can
easily be understood from the interference effects in the sums in (\ref{3.17})
and (\ref{3.20}).

\section{Bulk intermediate scattering functions}
The intermediate scattering function in the bulk follows by Fourier transform
with respect to the spatial coordinates. The incoherent part is obtained by
inserting (\ref{3.11}) and (\ref{3.14}) in (\ref{2.9}) with (\ref{2.3}):
\begin{eqnarray}
\fl S_{inc}({\bf k},t)=B^{-3/2}\; \int d(\brho-\brho') \; 
\rme^{-\rmi {\bkappa\cdot(\brho-\brho')}}
\; G_{\mu}^\ast({\bf r},{\bf r}',t)\; G({\bf r},{\bf r}',t)\nonumber\\
\lo= -\frac{(1+\rmi)B^{3/2}}{32\pi^3 \tau\sin(\thalf\tau)} \;{\sum_n}'\;
\rme^{\rmi(n+\thalf)\tau}\; J_{\perp,n}(\bkappa_{\perp},\tau)\; 
J_{\parallel,n,\nu}(\kappa_z,\tau)
\label{4.1}
\end{eqnarray}
with $\bkappa=(\bkappa_{\perp},\kappa_z)={\bf k}/\sqrt{B}$ and
$\brho=(\brho_{\perp},\zeta)=\sqrt{B}\;{\bf r}$. The integrals are defined as
\begin{eqnarray}
\fl J_{\perp,n}(\bkappa_{\perp},\tau)=\int d\brho_{\perp}\;
\exp\left[\half\rho_{\perp}^2/(\rme^{-\rmi\tau}-1)
-\rmi\bkappa_{\perp}\cdot\brho_{\perp}\right]\;
L_n\left(\half\rho_{\perp}^2\right)
\label{4.2}\\
\fl J_{\parallel,n,\nu}(\kappa_z,\tau)=\int d\zeta\; \rme^{-\rmi\kappa_z\zeta}
F^{\ast}\left[\sqrt{\tau(\nu-n-\thalf)},\frac{\zeta}{\sqrt{2\tau}}\right]
\label{4.3}
\end{eqnarray}
To evaluate $J_{\perp,n}$, we first perform the angular integral in terms of
a Bessel function. Subsequently, we use the identity \cite{ERD:1954}:
\begin{equation}
\fl
\int_0^{\infty}dx\; x\; \rme^{-\thalf a x^2}\; J_0(xy)\; L_n\left(\half
x^2\right)=
\frac{(a-1)^n}{a^{n+1}}\; \rme^{-\thalf y^2/a}\;
L_n\left(\frac{y^2}{2a(1-a)}\right)
\label{4.4}
\end{equation}
with $\Real a>0$ and $y>0$. In this way we get
\begin{equation}
\fl J_{\perp,n}(\bkappa_{\perp},\tau)=
4\pi\rmi\; \rme^{-\rmi(n+\thalf)\tau}\;\sin(\half\tau)\;
\exp\left[-\half\left(1-\rme^{-\rmi\tau}\right)\kappa_{\perp}^2\right]\;
L_n\left[2\sin^2(\half\tau)\kappa_{\perp}^2\right]
\label{4.5}
\end{equation}
For the integral over the parallel coordinate we employ the identity
\begin{equation}
\int_{-\infty}^{\infty} 
dz\; \rme^{-\rmi kz}\; F(a,zb)=\frac{2(1-\rmi)}{k}\; \rme^{\tfourth\rmi
k^2/b^2}\; \sin\left(\frac{ka}{|b|}\right)
\label{4.6}
\end{equation}
for real $k$, $a$ and $b$, which follows by a partial integration. In this way
we find
\begin{equation}
J_{\parallel,n,\nu}(\kappa_z,\tau)=\frac{2(1+\rmi)}{\kappa_z}\;
\rme^{-\thalf\rmi \kappa_z^2\tau}\;
\sin\left[\sqrt{2(\nu-n-\thalf)}\kappa_z\tau\right]
\label{4.7}
\end{equation}
Combining the above results, we have found the incoherent intermediate scattering
function as
\begin{eqnarray}
\fl S_{inc}({\bf k},t)=\frac{B^{3/2}}{2\pi^2\kappa_z\tau}\; 
\exp\left[-\half\rmi\kappa_z^2\tau
-\half\left(1-\rme^{-\rmi\tau}\right)\kappa_{\perp}^2\right]\nonumber\\
\times{\sum_n}'\sin\left[\sqrt{2(\nu-n-\thalf)}\kappa_z\tau\right]\; 
L_n\left[2\sin^2(\half\tau)\kappa_{\perp}^2\right]
\label{4.8}
\end{eqnarray}
For the special cases of purely transverse or purely parallel wavevectors the
incoherent intermediate scattering function can be simplified. For $\kappa_z=0$
it becomes an undamped periodic function of $\tau$ with period $2\pi$, whereas for
$\kappa_{\perp}=0$ it is a decaying function as in the general case. 

For $t=0$ the scattering function reduces to 
\begin{equation}
S_{inc}({\bf k},t=0)=\frac{B^{3/2}}{2\pi^2}\; {\sum_n}'
\sqrt{2(\nu-n-\thalf)}
\label{4.9}
\end{equation}
which is equal to the bulk particle density $n_{\infty}$ at the chosen chemical
potential. The time derivative at $t=0$ satisfies the sum rule
\begin{equation}
\left.\frac{\partial}{\partial t} S_{inc}({\bf k},t)\right|_{t=0}
=-\half\rmi\; k^2\;n_{\infty}
\label{4.10}
\end{equation}

We now turn to the coherent intermediate scattering function. By substituting 
(\ref{3.11}) in (\ref{2.9}) with (\ref{2.4}), we get
\begin{eqnarray}
\fl S_{coh}({\bf k},t)=-B^{-3/2}\; \int d(\brho-\brho') \; 
\rme^{-\rmi {\bkappa\cdot(\brho-\brho')}}
\; \left|G_{\mu}({\bf r},{\bf r}',t)\right|^2 \nonumber\\
\lo= -\frac{B^{3/2}}{16\pi^3 \tau} \;{\sum_m}'{\sum_n}'\;
\rme^{\rmi(m-n)\tau}\; J_{\perp,mn}(\bkappa_{\perp})\; 
J_{\parallel,mn,\nu}(\kappa_z,\tau)
\label{4.11}
\end{eqnarray}
The integrals are defined as
\begin{eqnarray}
\fl J_{\perp,mn}(\bkappa_{\perp})=\int d\brho_{\perp}\;
\rme^{-\thalf\rho_{\perp}^2
-\rmi\bkappa_{\perp}\cdot\brho_{\perp}}\;
L_m\left(\half\rho_{\perp}^2\right)\;
L_n\left(\half\rho_{\perp}^2\right)\label{4.12}\\
\fl J_{\parallel,mn,\nu}(\kappa_z,\tau)=\int d\zeta\; \rme^{-\rmi\kappa_z\zeta}
F^{\ast}\left[\sqrt{\tau(\nu-m-\thalf)},\frac{\zeta}{\sqrt{2\tau}}\right]
F\left[\sqrt{\tau(\nu-n-\thalf)},\frac{\zeta}{\sqrt{2\tau}}\right]
\label{4.13}
\end{eqnarray}
The angular integral in $J_{\perp,mn}$ leads to a Bessel function as
before. The radial integral can be determined by using the identity \cite{ERD:1954}
for $v>0$:
\begin{eqnarray}
\fl
\int_0^{\infty}du\; u\; \rme^{-\thalf u^2}\; J_0(uv)\; 
L_m\left(\half u^2\right)\; L_n\left(\half u^2\right)\nonumber\\
\lo= (-1)^{m+n}\; \rme^{-\thalf v^2}\; 
L_m^{-m+n}\left(\half v^2\right)\;L_n^{-n+m}\left(\half v^2\right)
\label{4.14}
\end{eqnarray}
which yields
\begin{equation}
\fl J_{\perp,mn}(\bkappa_{\perp})=
2\pi(-1)^{m+n}\; \rme^{-\thalf\kappa_{\perp}^2}\;
L_m^{-m+n}\left(\half \kappa_{\perp}^2\right)\;L_n^{-n+m}
\left(\half \kappa_{\perp}^2\right)
\label{4.15}
\end{equation}
The two Laguerre polynomials are not independent, as they are related as
\cite{MOS:1966}
\begin{equation}
(-1)^m\; m!\; z^{-m}\; L_m^{-m+n}(z)=(-1)^n\; n!\; z^{-n}\; L_n^{-n+m}(z)
\label{4.16}
\end{equation}
Hence, $J_{\perp,mn}(\bkappa_{\perp})$ is positive for all $m,n\geq 0$.
For the integral over the parallel coordinate we use the integral identity
(\ref{A.1}) from the appendix. With its help one gets
\begin{eqnarray}
\fl J_{\parallel,mn,\nu}(\kappa_z,\tau)=\nonumber\\
\fl =-\frac{2\rmi}{\kappa_z}\; \rme^{\thalf\rmi\kappa_z^2\tau}\;
\left\{\rme^{\rmi\kappa_z w_n\tau}\;
\theta\left[w_m^2-\left(\kappa_z+w_n\right)^2\right]
-\rme^{-\rmi\kappa_z w_n\tau}\;
\theta\left[w_m^2-\left(\kappa_z-w_n\right)^2\right]
\right\}\nonumber\\
\fl +\frac{2\rmi}{\kappa_z}\; \rme^{-\thalf\rmi\kappa_z^2\tau}\;
\left\{\rme^{-\rmi\kappa_z w_m\tau}\;
\theta\left[w_n^2-\left(\kappa_z+w_m\right)^2\right]
-\rme^{\rmi\kappa_z w_m\tau}\;
\theta\left[ w_n^2-\left(\kappa_z-w_m\right)^2\right]
\right\}
\label{4.17}
\end{eqnarray}
with the abbreviation $w_j=\sqrt{2(\nu-j-\thalf)}$. 

Substituting (\ref{4.15}) and (\ref{4.17}) in (\ref{4.11}) and employing the
symmetry with respect to an interchange of the summations over $m$ and $n$, we
obtain the coherent intermediate scattering function in the form
\begin{eqnarray}
\fl S_{coh}({\bf k},t)=\frac{B^{3/2}}{2\pi^2\kappa_z\tau}\; 
\rme^{-\thalf \kappa_{\perp}^2}\;{\sum_m}'{\sum_n}'\;
(-1)^{m+n}\;
L_m^{-m+n}\left(\half \kappa_{\perp}^2\right)\;L_n^{-n+m}
\left(\half \kappa_{\perp}^2\right)\nonumber\\
\times\left\{
\sin\left[\left(\half \kappa_z^2-\kappa_z w_m+n-m\right)\tau\right]\;
\theta\left[w_n^2-\left(\kappa_z-w_m\right)^2\right]\right.\nonumber\\
\left. -\sin\left[\left(\half \kappa_z^2+\kappa_z w_m+n-m\right)\tau\right]\;
\theta\left[w_n^2-\left(\kappa_z+w_m\right)^2\right]\right\}
\label{4.18}
\end{eqnarray}

For vanishing $\kappa_{\perp}$ the product of the Laguerre polynomials equals
$\delta_{mn}$. As a consequence, the coherent intermediate scattering function
gets the simpler form:
\begin{equation}
\fl S_{coh}({\bf k},t)=\frac{B^{3/2}}{2\pi^2|\kappa_z|\tau}\;
{\sum_n}'\; \sin\left[\left(\half \kappa_z^2-|\kappa_z| w_m\right)\tau\right]\;
\theta\left(2w_n-|\kappa_z|\right)
\label{4.19}
\end{equation}
A second special case arises for $\kappa_z=0$. Expanding the factor between
curly brackets in (\ref{4.18}) for small $\kappa_z$, one finds:
\begin{eqnarray}
\fl S_{coh}({\bf k},t)=-\frac{B^{3/2}}{2\pi^2}\;
\rme^{-\thalf \kappa_{\perp}^2}\;{\sum_m}'{\sum_n}'\;
(-1)^{m+n}\;
L_m^{-m+n}\left(\half \kappa_{\perp}^2\right)\;L_n^{-n+m}
\left(\half \kappa_{\perp}^2\right) \nonumber\\
\fl\times\left\{\cos[(m-n)\tau]\;\left[(1-\delta_{mn})\; \theta(m-n)\; w_m
+(1-\delta_{mn})\; \theta(n-m)\; w_n\right]
+\delta_{mn}\; w_m\right\}
\label{4.20}
\end{eqnarray}
In this case the coherent intermediate scattering function is an undamped periodic
function of $\tau$.

The static function follows by putting $t=0$ in (\ref{4.18}). Rearranging the
terms one may rewrite it as
\begin{eqnarray}
\fl S_{coh}({\bf k},t=0)=-\frac{B^{3/2}}{4\pi^2}\; 
\rme^{-\thalf \kappa_{\perp}^2}\;{\sum_m}'{\sum_n}'\;
(-1)^{m+n}\;
L_m^{-m+n}\left(\half \kappa_{\perp}^2\right)\;L_n^{-n+m}
\left(\half \kappa_{\perp}^2\right)\nonumber\\
\fl\times \left[
\left(w_m+w_n-|\kappa_z|\right)\;\theta\left(w_m+w_n-|\kappa_z|\right)
-\left(w_m-w_n-|\kappa_z|\right)\;\theta\left(w_m-w_n-|\kappa_z|\right)\right.
\nonumber\\
\left.-\left(w_n-w_m-|\kappa_z|\right)\;\theta\left(w_n-w_m-|\kappa_z|\right)\right]
\label{4.21}
\end{eqnarray}
In contrast to $S_{inc}$, the coherent intermediate scattering function is an
even function of $t$, as is clear from (\ref{4.18}). Hence, the time derivative
of $S_{coh}$ at $t=0$ vanishes, whereas the time derivative of $S_{inc}$ at
$t=0$ is different from 0, as we have seen in (\ref{4.10}).

On comparing (\ref{4.8}) and (\ref{4.18}) one notes that the two parts of the
intermediate scattering function are given by rather different
expressions. However, they can be made more analogous by employing the identity
\begin{eqnarray}
\fl\sum_{m=0}^{\infty}(-1)^m\; \rme^{-\rmi(m+\thalf)\tau}\;
L_m^{-m+n}\left(\half \kappa_{\perp}^2\right)\;L_n^{-n+m}
\left(\half \kappa_{\perp}^2\right)\nonumber\\
\lo= (-1)^n\;\rme^{-\rmi(n+\thalf)\tau}\;
L_n\left[2\sin^2\left(\half\tau\right)\;\kappa_{\perp}^2\right]\;
\exp\left(\half \rme^{-\rmi\tau}\;\kappa_{\perp}^2\right)
\label{4.22}
\end{eqnarray}
which may be proved by using (\ref{4.14}) in reverse order, and then
(\ref{3.13}) and (\ref{4.4}). In this way we may write (\ref{4.8}) as
\begin{eqnarray}
\fl S_{inc}({\bf k},t)=\frac{B^{3/2}}{2\pi^2\kappa_z\tau}\; 
\rme^{-\thalf \kappa_{\perp}^2-\thalf\rmi\kappa_z^2\tau}
\sum_{m=0}^{\infty}{\sum_n}'\;
(-1)^{m+n}\; \rme^{\rmi(n-m)\tau}\nonumber\\
\times L_m^{-m+n}\left(\half \kappa_{\perp}^2\right)\;
L_n^{-n+m}\left(\half \kappa_{\perp}^2\right)\;
\sin(w_n\kappa_z \tau)\
\label{4.23}
\end{eqnarray}
This result could have been obtained in an alternative way by substituting
(\ref{3.12}) (instead of (\ref{3.14})) in (\ref{2.9}) with (\ref{2.3}), and
taking similar steps as in (\ref{4.11})--(\ref{4.18}). 

\section{Bulk dynamical structure factor}
The dynamical structure factor follows from the intermediate scattering function
by taking a Fourier transform with respect $t$. For the incoherent part of the
structure factor we find from (\ref{4.23}):
\begin{eqnarray}
\fl S_{inc}({\bf k},\omega)=B^{-1}\int_{-\infty}^{\infty}d\tau\; 
\rme^{\rmi\varpi\tau}\; S_{inc}({\bf k},t)\nonumber\\
\fl =\frac{B^{1/2}}{2\pi|\kappa_z|}\; \rme^{-\thalf \kappa_{\perp}^2}\sum_{m=0}^{\infty}
{\sum_n}' (-1)^{m+n} \; L_m^{-m+n}\left(\half \kappa_{\perp}^2\right)\;
L_n^{-n+m}\left(\half \kappa_{\perp}^2\right)\;\theta\left[\nu-\nu_{mn}(\omega)\right]
\label{5.1}
\end{eqnarray}
where we introduced the rescaled frequency $\varpi=\omega/B$, and the abbreviation
\begin{equation}
\nu_{mn}(\omega)=n+\half+\half\kappa_z^{-2}\left(\half\kappa_z^2-\varpi+m-n\right)^2
\label{5.2}
\end{equation}
The dependence of $S_{inc}$ on $\omega$ is determined by the step function. For
negative $\omega$ it implies that the summation variable $m$ is confined to
values below $\nu-\half$, as is $n$. For positive $\omega$ this need not be the
case.

The coherent part of the dynamic structure factor follows from (\ref{4.18}) as
\begin{eqnarray}
\fl S_{coh}({\bf k},\omega)=-\frac{B^{1/2}}{2\pi|\kappa_z|}
\rme^{-\thalf \kappa_{\perp}^2}{\sum_m}'
{\sum_n}' (-1)^{m+n} \nonumber\\
\times L_m^{-m+n}\left(\half \kappa_{\perp}^2\right)
L_n^{-n+m}\left(\half \kappa_{\perp}^2\right)
\theta\left[\nu-\nu_{mn}(-|\omega|)\right]
\label{5.3}
\end{eqnarray}
Again the $\omega$-dependence shows up through the step function only.

\mbox{}From (\ref{4.16}) it follows that the incoherent part of the structure function
is positive for all ${\bf k}$ and $\omega$, whereas the coherent part is
negative. Furthermore, on comparing (\ref{5.1}) and (\ref{5.3}) one finds that
the two parts of the dynamic structure factor are connected as $S_{coh}({\bf
k},\omega)=-S_{inc}({\bf k},-|\omega|)$. This relation is in agreement with the
general identity (\ref{2.12}), since the structure factor is symmetric under 
spatial inversion. It implies that the total dynamic structure factor at $T=0$
vanishes for negative $\omega$.

If the wavevector is purely transverse, the expressions (\ref{5.1}) and
(\ref{5.3}) cannot be used as such. It is more convenient in this case to take
the limit $\kappa_z=0$ in (\ref{4.23}) and to use (\ref{4.20}) before taking the
Fourier transform. In this way one gets
\begin{eqnarray}
\fl S_{inc}({\bf k},\omega)=\frac{B^{1/2}}{\pi}\; 
\rme^{-\thalf \kappa_{\perp}^2}
\sum_{m=0}^{\infty}{\sum_n}'\;(-1)^{m+n}\nonumber\\
\times L_m^{-m+n}\left(\half \kappa_{\perp}^2\right)\;
L_n^{-n+m}\left(\half \kappa_{\perp}^2\right)\;
w_n \; \delta(\varpi -m+n)
\label{5.4}
\end{eqnarray}
and $S_{coh}({\bf k},\omega)=-S_{inc}({\bf k},-|\omega|)$, as before.  Both
parts of the dynamical structure factor are sums over delta functions, with
peaks at the integer resonance frequencies $m-n$.  

For wavevectors parallel to the field one finds from (\ref{5.1}):
\begin{equation}
S_{inc}({\bf k},\omega)=\frac{B^{1/2}}{2\pi |\kappa_z|}\; 
{\sum_n}'\; \theta\left[\nu-\nu_{nn}(\omega)\right]
\label{5.5}
\end{equation}
and an analogous expression for the coherent part. The sum of step functions
reaches its maximum value for $\varpi=\half\kappa_z^2$.

\begin{figure}
  \begin{center}
    \includegraphics[height=3.5cm]{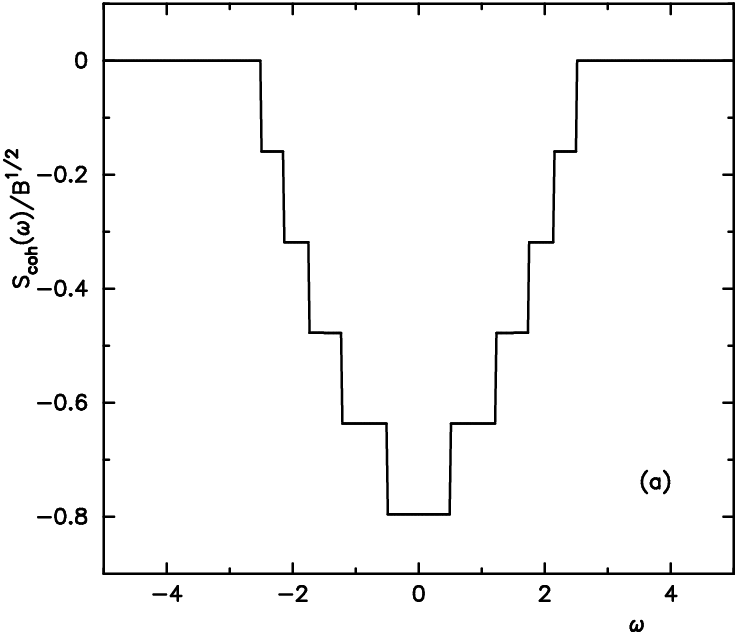}
    \hspace{0.1cm}
    \includegraphics[height=3.5cm]{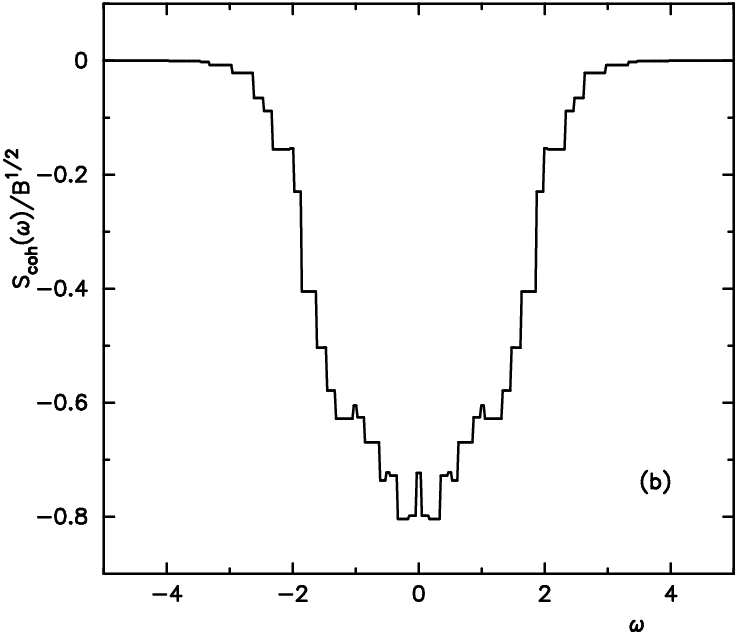}
    \hspace{0.1cm}
    \includegraphics[height=3.5cm]{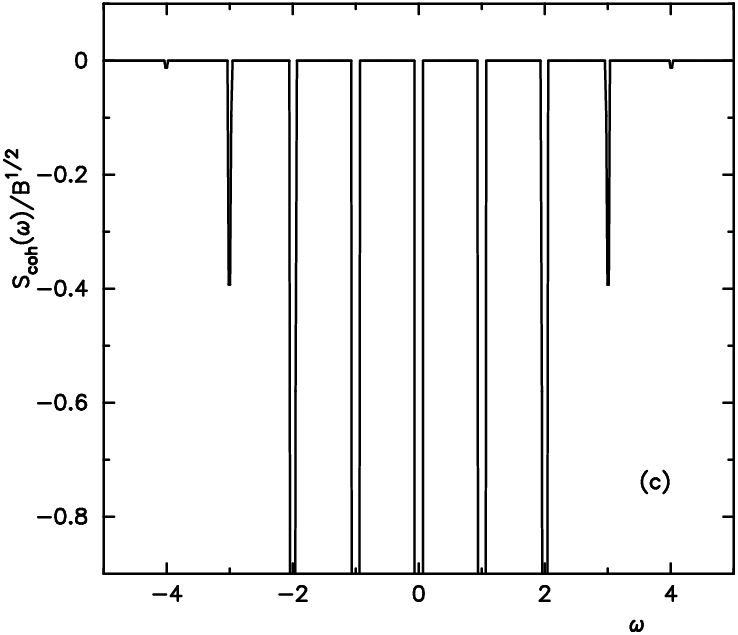}
  \caption{Bulk coherent structure factor $S_{coh}({\bf k},\omega)/B^{1/2}$ for
    $\nu=5$ and $|\bkappa|=1$, as a function of $\varpi$. The curves represent
    the structure factor for $\theta=0$ (a), $\theta=\pi/4$ (b), and
    $\theta=1.55$ (c), with $\theta=\arctan(|\bkappa_{\perp}|/\kappa_z)$.}
\label{fig2}
  \end{center}
\end{figure}

\begin{figure}
  \begin{center}
    \includegraphics[height=3.5cm]{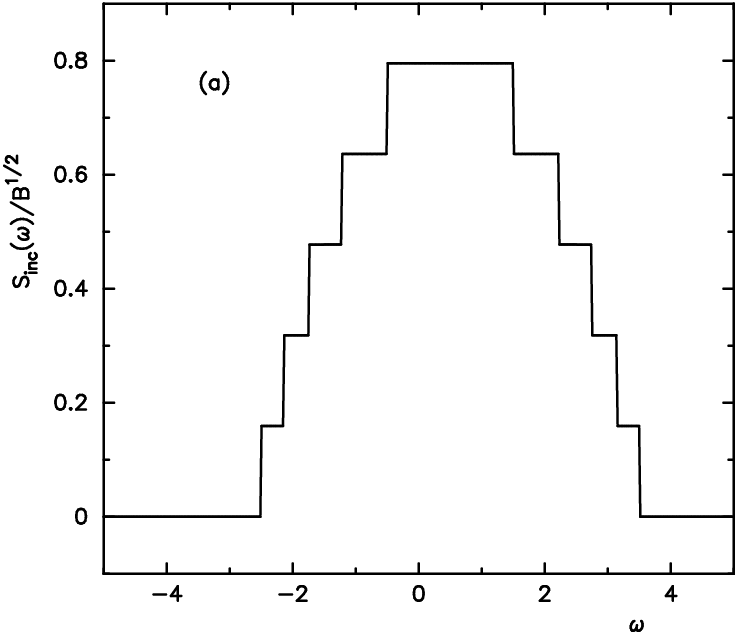}
    \hspace{0.1cm}
    \includegraphics[height=3.5cm]{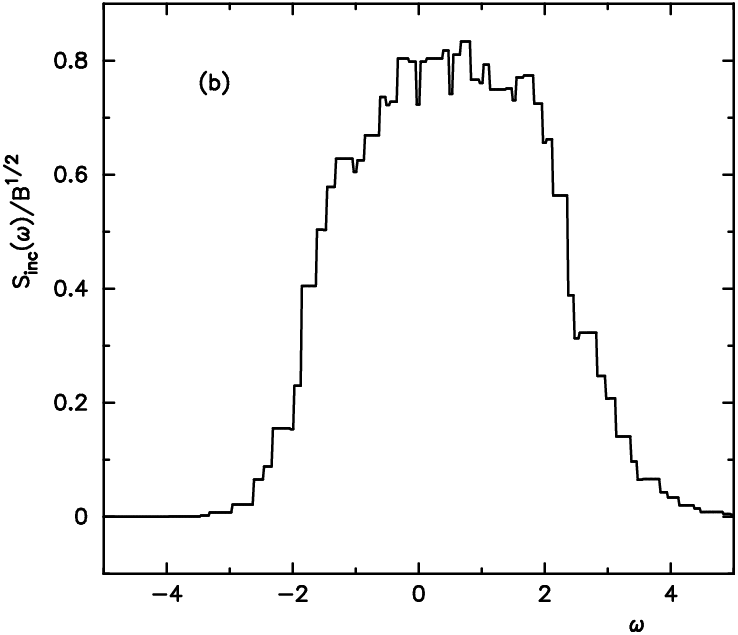}
    \hspace{0.1cm}
    \includegraphics[height=3.5cm]{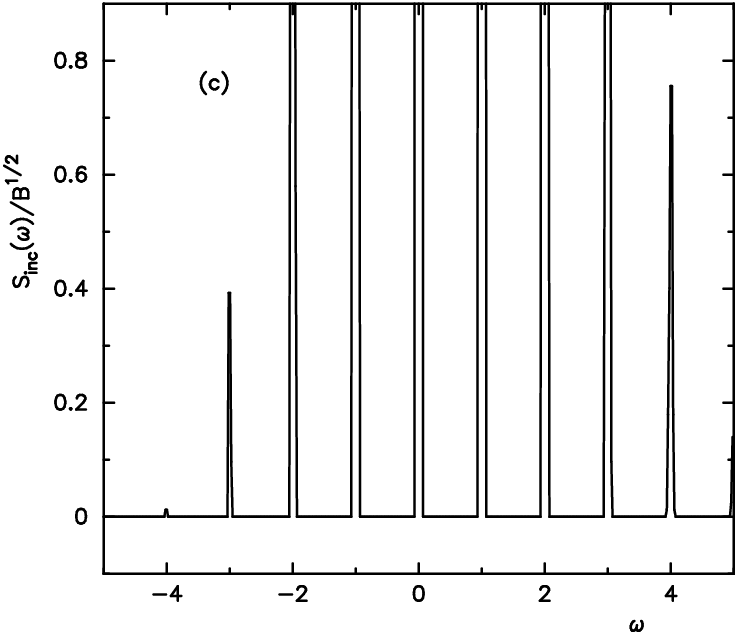}
  \caption{Bulk incoherent structure factor $S_{inc}({\bf k},\omega)/B^{1/2}$ for
    $\nu=5$ and $|\bkappa|=1$, as a function of $\varpi$. The curves represent
    the structure factor for $\theta=0$ (a), $\theta=\pi/4$ (b), and
    $\theta=1.55$ (c), with $\theta=\arctan(|\bkappa_{\perp}|/\kappa_z)$.}
\label{fig3}
  \end{center}
\end{figure}

The numerical results in figure \ref{fig2} show how the coherent dynamic
structure factor changes from a function with steps for a purely longitudinal
wavevector to a set of discrete spikes for a (nearly) transverse wavevector. A
similar behavior is found for the incoherent dynamic structure factor, as shown
in figure \ref{fig3}. For this case the curves are asymmetric with respect to a
change of sign of $\omega$.

The sum rules (\ref{4.9}) and (\ref{4.10}) for the incoherent intermediate
scattering function yield
\begin{equation}
\int_{-\infty}^{\infty}d\omega \; S_{inc}({\bf k},\omega)=2\pi n_{\infty}
\quad , \quad 
\int_{-\infty}^{\infty}d\omega \; \omega S_{inc}({\bf k},\omega)=\pi k^2
n_{\infty}
\label{5.6}
\end{equation}

The above results for the bulk dynamical structure factor at $T=0$ can easily be
generalized to finite temperatures. In fact, for the incoherent part
(\ref{2.11}), one may use an identity analogous to (\ref{2.8}) to obtain the
expression for finite $T$ from that for $T=0$. Since the chemical potential in
(\ref{5.1}) only occurs in the step function, the integral over $\mu'$ is
trivial. In this way one easily arrives at the result for finite temperature:
\begin{eqnarray}
\fl S_{inc}({\bf k},\omega)=
\frac{B^{1/2}}{2\pi|\kappa_z|}\; \rme^{-\thalf \kappa_{\perp}^2}\sum_{m=0}^{\infty}
\sum_{n=0}^{\infty} (-1)^{m+n} \nonumber\\
\times L_m^{-m+n}\left(\half \kappa_{\perp}^2\right)\;
L_n^{-n+m}\left(\half
\kappa_{\perp}^2\right)\;\frac{1}{\rme^{\beta[\nu_{mn}(\omega)-\nu]}+1}
\label{5.7}
\end{eqnarray}
where the temperature is given in units of $B$. The coherent part of the dynamic
structure factor for finite temperatures follows from (\ref{2.12}):
\begin{eqnarray}
\fl S_{coh}({\bf k},\omega)=
\frac{B^{1/2}}{2\pi|\kappa_z|}\; \rme^{-\thalf \kappa_{\perp}^2}\sum_{m=0}^{\infty}
\sum_{n=0}^{\infty} (-1)^{m+n}\; L_m^{-m+n}\left(\half \kappa_{\perp}^2\right)\;
L_n^{-n+m}\left(\half\kappa_{\perp}^2\right)\nonumber\\
\times\left\{
\frac{1}{\rme^{\beta\omega}-1}\;
\frac{1}{\rme^{\beta[\nu_{mn}(\omega)-\nu]}+1}+
\frac{1}{\rme^{-\beta\omega}-1}\;
\frac{1}{\rme^{\beta[\nu_{mn}(-\omega)-\nu]}+1}\right\}
\label{5.8}
\end{eqnarray}

\section{Vanishing magnetic field}
In the limit of vanishing magnetic field the results of the previous sections
are expected to reduce to the well-known expressions for a non-magnetized
degenerate free-electron gas. We may check this by considering the limiting
behavior of the functions $F_{\mu}$ en $G$ in position space. 

The expression (\ref{3.11}) for $G_{\mu}$ contains a sum that extends to
infinity for $B\rightarrow 0$. Hence, the order $n$ of the Laguerre polynomial
may become large. On the other hand, its argument gets small, since the position
variables scale with $\sqrt{B}$. In this regime the Laguerre polynomials may be
approximated as \cite{ERD:1953}:
\begin{equation}
L_n(u)\approx \rme^{\thalf u}\; J_0\left(\sqrt{2(2n+1)u}\right)
\label{6.1}
\end{equation}
Putting $nB=v$, we may write the sum in (\ref{3.11}) 
as an integral for small
$B$. Replacing the scaled variables by the original ones, we get for $B=0$:
\begin{eqnarray}
\fl G_{\mu}({\bf r},{\bf r}',t)=\frac{1}{4\pi^{3/2}t^{1/2}}\; 
\rme^{\thalf\rmi(z-z')^2/t}
\; \int_0^{\mu}dv\; \rme^{-\rmi vt}\; 
J_0\left(\sqrt{2v}\; |{\bf r}_{\perp}-{\bf r}_{\perp}'|\right) \nonumber\\
\times F\left(\sqrt{t(\mu-v)},\frac{z-z'}{\sqrt{2t}}\right)
\label{6.2}
\end{eqnarray}
With the use of the relation (\ref{A.7}) of the appendix we obtain
\begin{eqnarray}
\fl G_{\mu}({\bf r},{\bf r}',t)=
-\frac{\rmi}{4\pi^{3/2}t^{3/2}}\; \rme^{\thalf\rmi|{\bf
r}-{\bf r}'|^2/t}\; F\left(\sqrt{\mu t},\frac{|{\bf r}-{\bf
r}'|}{\sqrt{2t}}\right)\nonumber\\
+\frac{\rmi}{2\pi^2|{\bf r}-{\bf r}'|t}\;
\rme^{-\rmi \mu t}\; \sin\left(\sqrt{2\mu}\; |{\bf r}-{\bf r}'|\right)
\label{6.3}
\end{eqnarray}
for $B=0$. For large $t$ the second term dominates. It factorizes in a
space-dependent and a time-dependent factor. The latter is proportional to
$t^{-1}$ (apart from a $t$-dependent phase factor). This is in contrast to the
behavior in the case with field. For ${\bf r}={\bf r}'$ we have seen in
(\ref{3.24}) that $G_{\mu}$ has a decay proportional to $t^{-1/2}$ times the ratio
of two time-dependent sine functions.

The limiting form of $G$ for vanishing $B$ follows immediately from
(\ref{3.14}). One finds
\begin{equation}
G({\bf r},{\bf r}',t)=-\frac{1+\rmi}{4\pi^{3/2}t^{3/2}}\; \rme^{\thalf\rmi|{\bf
r}-{\bf r}'|^2/t}
\label{6.4}
\end{equation}
It decays algebraically in time, and is free from the singularities that
characterize the expression (\ref{3.14}) for the case with field. For small
fields the poles in (\ref{3.14}) at $\tau=2\pi m$, or $t=2\pi m/B$, with integer
$m\neq 0$, shift towards $\infty$, and disappear for vanishing field. Only the
singularity at $t=0$ remains. It is connected to the delta function singularity
of the static function $G({\bf r},{\bf r}',t=0)$.

The incoherent and coherent parts of the dynamic structure factor follow by
taking Fourier transforms, as we have seen in the previous sections. As a useful
check of these results we may verify that the limits of (\ref{5.1}) and
(\ref{5.3}) for vanishing field agree with the well-known results for a
field-free degenerate fermion gas.  Starting with the incoherent part
(\ref{5.1}), we first use (\ref{4.14}) in reverse order. The resulting integral
over $u$, with $v=\kappa_{\perp}$, contains a Bessel function
$J_0(u\kappa_{\perp})$. For small field, $\kappa_{\perp}$ becomes large so that
effectively only small values of $u$ contribute in the integral. Hence, one may
use (\ref{6.1}) for the Laguerre polynomials in the integral over $u$, so that
the latter becomes an integral over the product of three Bessel functions. It
can be evaluated as a product of two inverse square roots \cite{ERD:1954}. In
this way we have found for large $v$, and large $m,n$:
\begin{eqnarray}
\fl (-1)^{m+n}\; \rme^{-\thalf v^2}\; 
L_m^{-m+n}\left(\half v^2\right)\;L_n^{-n+m}\left(\half v^2\right)\nonumber\\
\lo{\approx} \frac{2}{\pi\sqrt{16mn-(v^2-2m-2n)^2}}\; 
\theta\left[16mn-(v^2-2m-2n)^2\right]
\label{6.5}
\end{eqnarray}
Employing this asymptotic relation in (\ref{5.1}) and replacing the sums over
$m,n$ with integrals, by writing $x=mB$ and $y=nB$, we get for $B=0$
\begin{eqnarray}
\fl S_{inc}({\bf k},\omega)
=\frac{1}{2\pi^2|k_z|}\;\int_0^{\infty}dx\int_0^{\mu}dy\;
\frac{1}{\sqrt{-x^2+2xy-y^2+k_{\perp}^2(x+y)-\tfourth k_{\perp}^4}}\nonumber\\
\fl\times\theta\left[-x^2+2xy-y^2+k_{\perp}^2(x+y)-
\tfourth k_{\perp}^4\right]\nonumber\\
\fl\times\theta\left[-x^2+2xy-y^2-k_z^2(x+y)+2\omega(x-y)-
(\thalf k_z^2-\omega)^2+2\mu k_z^2\right]
\label{6.6}
\end{eqnarray}
The two step functions determine a region in the $xy$-plane that is enclosed
between two parabolas. Integrating first over $x+y$, and subsequently over
$x-y$, we arrive at the expression
\begin{equation}
S_{inc}({\bf k},\omega)=\frac{\mu}{2\pi k}\;
\left[1-\half\left(\frac{\omega}{\sqrt{\mu}\; k}-
\frac{k}{2\sqrt{\mu}}\right)^2\right]
\label{6.7}
\end{equation}
valid for $-\sqrt{2\mu}\; k+\half k^2\leq \omega\leq \sqrt{2\mu}\; k+\half k^2$.
The coherent part of the dynamic structure factor may be discussed along similar
lines. One finds in the field-free limit
\begin{equation}
S_{coh}({\bf k},\omega)=-\frac{\mu}{2\pi k}\;
\left[1-\half\left(\frac{|\omega|}{\sqrt{\mu}\; k}+
\frac{k}{2\sqrt{\mu}}\right)^2\right]
\label{6.8}
\end{equation}
for $k\leq 2\sqrt{2\mu}$ and $|\omega|\leq \sqrt{2\mu}\; k-\half k^2$. The
expressions (\ref{6.7}) and (\ref{6.8}) agree with the well-known results for a
degenerate fermion gas without a magnetic field \cite{LIN:1954},
\cite{PIN:1966}, \cite{MAZ:1996}.

\section{Edge effects: hard wall perpendicular to the field}
In the presence of a hard wall at $z=0$ the energy eigenfunctions
$\varphi_{n_y,n_z,n}({\bf r})$ in the domain $z>0$ are related to (\ref{3.2}) by
a reflection principle:
\begin{equation}
\varphi_{n_y,n_z,n}({\bf r})=\frac{1}{\sqrt{2}}\left[\psi_{n_y,n_z,n}({\bf r})-
\psi_{n_y,n_z,n}(\tilde{\bf r})\right]
\label{7.1}
\end{equation}
with $\tilde{\bf r}=(x,y,-z)$. Hence, both $G_{\mu}$ and $G$ can be obtained
immediately from the results of section 3. We will only consider the case
$z=z'$. From (\ref{3.11}) we get:
\begin{eqnarray}
\fl G_{\mu}({\bf r},{\bf r}',t)=
\frac{B^{3/2}}{4\pi^{3/2} \tau^{1/2}} \;
\rme^{-\tfourth(\brho_{\perp}-\brho_{\perp}')^2
+\thalf\rmi(\xi+\xi')(\eta-\eta')}{\sum_n}' \rme^{-\rmi(n+\thalf)\tau}\;
 L_n\left[\half(\brho_{\perp}-\brho_{\perp}')^2\right] \nonumber\\ 
\times\; 
\left\{ F\left[\sqrt{\tau(\nu-n-\thalf)},0\right]-
\rme^{2\rmi\zeta^2/\tau}
F\left[\sqrt{\tau(\nu-n-\thalf)},\sqrt{\frac{2}{\tau}}\zeta\right]\right\}
\label{7.2} 
\end{eqnarray}
Likewise, we find from (\ref{3.14}):
\begin{equation}
\fl G({\bf r},{\bf r}',t)= -\frac{(1+\rmi)B^{3/2}}{8\pi^{3/2}
\tau^{1/2}\sin(\thalf\tau)}\;\left(1-\rme^{2\rmi\zeta^2/\tau}\right) 
\;\rme^{\tfourth\rmi\cot(\thalf\tau)(\brho_{\perp}-\brho_{\perp}')^2
+\thalf\rmi(\xi+\xi')(\eta-\eta')}\;
\label{7.3}
\end{equation}
For equal position vectors (${\bf r}={\bf r}'$) the Laguerre polynomial and
the space-dependent exponential drop out from (\ref{7.2}), so that one gets:
\begin{eqnarray}
\fl G_{\mu}({\bf r},{\bf r},t)=
\frac{B^{3/2}}{4\pi^{3/2} \tau^{1/2}} \;
{\sum_n}' \rme^{-\rmi(n+\thalf)\tau}\;
 \nonumber\\ 
\times\; 
\left\{ F\left[\sqrt{\tau(\nu-n-\thalf)},0\right]-
\rme^{2\rmi\zeta^2/\tau}
F\left[\sqrt{\tau(\nu-n-\thalf)},\sqrt{\frac{2}{\tau}}\zeta\right]\right\}
\label{7.4} 
\end{eqnarray}
The expression (\ref{7.3}) for $G$ reduces in this case to
\begin{equation}
G({\bf r},{\bf r},t)= -\frac{(1+\rmi)B^{3/2}}{8\pi^{3/2}
\tau^{1/2}\sin(\thalf\tau)}\;\left(1-\rme^{2\rmi\zeta^2/\tau}\right) 
\label{7.5}
\end{equation}

The static correlation function $G_{\mu}(t=0)$ follows upon using (\ref{3.15})
in (\ref{7.2}):
\begin{eqnarray}
\fl G_{\mu}({\bf r},{\bf r}',t=0)=\frac{B^{3/2}}{2\pi^2}\;
\rme^{-\tfourth(\brho_{\perp}-\brho_{\perp}')^2
+\thalf\rmi(\xi+\xi')(\eta-\eta')}\;{\sum_n}'\; 
L_n\left[\half(\brho_{\perp}-\brho_{\perp}')^2\right]\nonumber\\
\times\sqrt{2(\nu-n-\thalf)}
\left\{1-\frac{\sin\left[2\sqrt{2(\nu-n-\thalf)}\;\zeta\right]}
{2\sqrt{2(\nu-n-\thalf)}\;\zeta}\right\}
\label{7.6}
\end{eqnarray}
This expression may also be obtained directly from (\ref{3.16}). For coinciding
positions the Laguerre polynomial and the exponential prefactor again drop
out. The function $G$ reduces to $\delta({\bf r}-{\bf r}')$ in the static limit,
as in the bulk case.

The expressions (\ref{7.2}) and (\ref{7.3}) (or (\ref{7.4}) and
(\ref{7.5}) for coinciding positions) for $G_{\mu}$ and $G$ may be compared to their
counterparts (\ref{3.11}) and (\ref{3.14}) (or (\ref{3.22}) and (\ref{3.23}))
for the bulk. The contributions from the reflection in the wall lead to a change
in the dependence of the correlation function on time. This becomes manifest
upon choosing positions near the wall. A Taylor expansion for small $\zeta$ 
leads to the following expressions for the correlation functions near the wall:
\begin{eqnarray}
\fl G_{\mu}({\bf r},{\bf r}',t)\approx
-\frac{\rmi \zeta^2 B^{3/2}}{\pi^{3/2} \tau^{3/2}} \;
\rme^{-\tfourth(\brho_{\perp}-\brho_{\perp}')^2
+\thalf\rmi(\xi+\xi')(\eta-\eta')} \nonumber\\
\fl\times {\sum_n}' L_n\left[\half(\brho_{\perp}-\brho_{\perp}')^2\right]\; \left\{
\rme^{-\rmi(n+\thalf)\tau}\;\left[ 
C\left(\sqrt{\tau(\nu-n-\thalf)}\right)\right.\right.\nonumber\\
\left.\left.-\rmi S\left(\sqrt{\tau(\nu-n-\thalf)}\right)\right]-
\sqrt{\frac{2}{\pi}}\; \rme^{-\rmi\nu\tau}\; \sqrt{\tau(\nu-n-\thalf)}\right\}
\label{7.7} 
\end{eqnarray}
and 
\begin{eqnarray}
\fl G({\bf r},{\bf r}',t)\approx -\frac{(1-\rmi)\zeta^2 B^{3/2}}{4\pi^{3/2}
\tau^{3/2}\sin(\thalf\tau)}\;
\rme^{\tfourth\rmi\cot(\thalf\tau)(\brho_{\perp}-\brho_{\perp}')^2
+\thalf\rmi(\xi+\xi')(\eta-\eta')}\nonumber\\
\mbox{}
\label{7.8}
\end{eqnarray}
For equal positions these become:
\begin{eqnarray}
\fl G_{\mu}({\bf r},{\bf r},t)\approx
-\frac{\rmi \zeta^2 B^{3/2}}{\pi^{3/2} \tau^{3/2}} \;
{\sum_n}' \left\{
\rme^{-\rmi(n+\thalf)\tau}\;\left[ 
C\left(\sqrt{\tau(\nu-n-\thalf)}\right)\right.\right.\nonumber\\
\left.\left.-\rmi S\left(\sqrt{\tau(\nu-n-\thalf)}\right)\right]
-\sqrt{\frac{2}{\pi}}\; \rme^{-\rmi\nu\tau}\; \sqrt{\tau(\nu-n-\thalf)}\right\}
\label{7.9} 
\end{eqnarray}
and 
\begin{equation}
G({\bf r},{\bf r},t)\approx -\frac{(1-\rmi)\zeta^2 B^{3/2}}{4\pi^{3/2}
\tau^{3/2}\sin(\thalf\tau)}\;
\label{7.10}
\end{equation}
For large $\tau$ the Fresnel integrals between the curly brackets in (\ref{7.7})
and (\ref{7.9}) can be neglected, so that both expressions show an algebraic
decay proportional to $\tau^{-1}$. From (\ref{7.7}) we get in this way
\begin{eqnarray}
\fl G_{\mu}({\bf r},{\bf r}',t)\approx
\frac{\rmi \zeta^2 B^{3/2}}{\pi^2 \tau} \;
\rme^{-\tfourth(\brho_{\perp}-\brho_{\perp}')^2
+\thalf\rmi(\xi+\xi')(\eta-\eta')-\rmi\nu\tau} \nonumber\\
\times {\sum_n}' L_n\left[\half(\brho_{\perp}-\brho_{\perp}')^2\right]\; 
\sqrt{2(\nu-n-\thalf)}
\label{7.11} 
\end{eqnarray}
which for coinciding positions reduces to:
\begin{equation}
\fl G_{\mu}({\bf r},{\bf r},t)\approx
\frac{\rmi \zeta^2 B^{3/2}}{\pi^{2} \tau} \;\rme^{-\rmi\nu\tau}\;
{\sum_n}'  \sqrt{2(\nu-n-\thalf)}=\frac{2\rmi\zeta^2}{\tau}\;
\rme^{-\rmi\nu\tau}\;
n_{\infty}
\label{7.12} 
\end{equation}
with $n_{\infty}$ the bulk particle density. 

Comparing (\ref{7.12}) with (\ref{3.24}), one concludes that the behavior of the
correlation function $G_{\mu}$, and hence of the coherent structure function
$S_{coh}$, has changed owing to the presence of the wall. In fact, whereas in
the bulk the coherent structure function for coinciding positions decays
proportional to $1/\tau$ for large time (with a periodic modulation), the decay
near the wall is faster, namely proportional to $\tau^{-2}$.  The function $G$
has changed as well, with a decay proportional to $\tau^{-3/2}$ near the wall,
which leads to a corresponding change in the behavior of $S_{inc}$. However, the
recurring singularities at $\tau=2\pi m$ dominate the picture completely in this
case, and these are not altered by the wall.

\begin{figure}
  \begin{center}
    \includegraphics[height=5.cm]{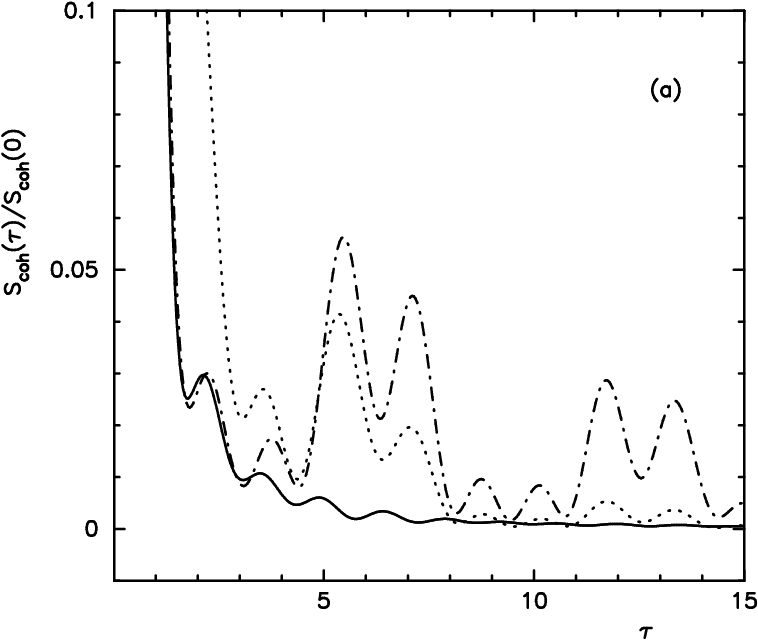}
    \hspace{0.5cm}
    \includegraphics[height=5.0cm]{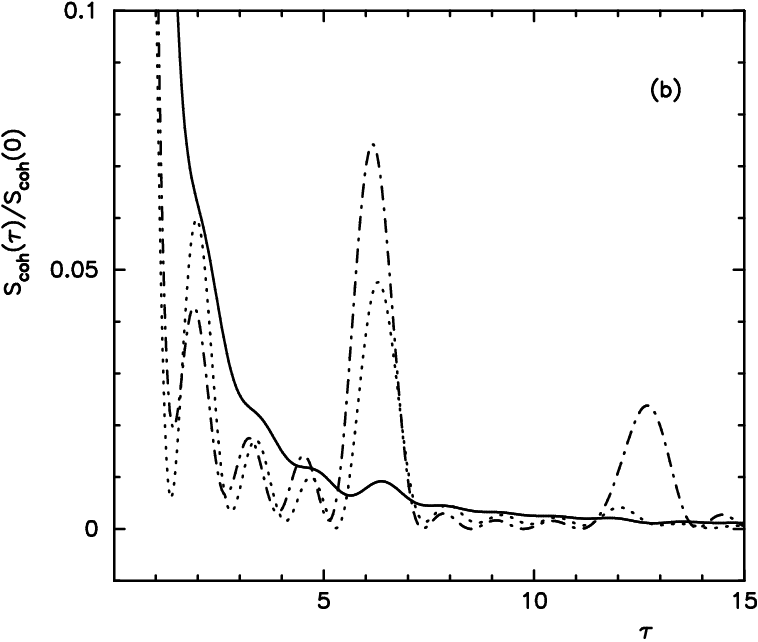}
  \caption{Coherent structure function as a function of time $\tau$, at
    $\nu=5$, for position difference $|\xi-\xi'|=1$ (a), and for coinciding
    positions (b). The curves, which start at $1$ for $\tau=0$, give the
    normalized structure function for distances $\zeta$ from the wall equal to
    $0.1$ (\full), $1.5$ (\dotted) and $\infty$ (bulk, \chain).}
\label{fig4}
  \end{center}
\end{figure}

In figure \ref{fig4} the normalized coherent structure function $S_{coh}({\bf
r},{\bf r}',t)/S_{coh}({\bf r},{\bf r}',0)$ is shown for several values of the
distance from the wall $\zeta$ and the position difference $|\xi-\xi'|$. The
faster time decay of the structure function for positions near the wall is
manifest. Revivals are (almost) completely suppressed in this case.

\section{Edge effects: hard wall parallel to the field}
Let us now consider wall effects in a different geometric arrangement, namely
for a wall parallel to the field.  For a hard wall at $x=0$ the energy
eigenfunctions $\varphi_{n_y,n_z,n}({\bf r})$ for $x>0$ can be expressed in parabolic
cylinder functions \cite{KES:2001}, \cite{ERD:1953} as
\begin{equation}
\fl \varphi_{n_y,n_z,n}({\bf r})=\frac{B^{1/4}}{L}\;
\frac{D_{\varepsilon_n(\kappa_y)-1/2}\left[\sqrt{2}(\xi-\kappa_y)\right]}
{\left\{\int_0^{\infty}d\xi'\; D^2_{\varepsilon_n(\kappa_y)-1/2}
\left[\sqrt{2}(\xi'-\kappa_y)\right]\right\}^{1/2}}\;
\rme^{\rmi\kappa_y\eta+\rmi\kappa_z\zeta}
\label{8.1}
\end{equation}
As for the bulk case, the wavevectors are given as $\kappa_i=2\pi
n_i/(L\sqrt{B})$, with integer $n_i$ for $i=y,z$. Furthermore, the function
$\varepsilon_n(\kappa)$ occurring in the index of the cylinder functions is
determined by \cite{KES:2001}
\begin{equation}
D_{\varepsilon_n(\kappa)-1/2}\left(-\sqrt{2}\,\kappa\right)=0
\label{8.2}
\end{equation}
The same function shows up in the eigenvalue associated to (\ref{8.1}) 
\begin{equation}
E_{n_y,n_z,n}=B\; \varepsilon_n(\kappa_y)+\half k_z^2
\label{8.3}
\end{equation}

The time correlation function $G_{\mu}$ for the degenerate case follows by
inserting (\ref{8.1}) in (\ref{2.7}). We will only
consider the case $x=x'$. Using (\ref{3.6}) and (\ref{3.9}), with $n+\half$
replaced by $\varepsilon_n(\kappa_y)$, we get
\begin{eqnarray}
\fl G_{\mu}({\bf r},{\bf r}',t)=\frac{B^{3/2}}{4\pi^{3/2}\tau^{1/2}}\;
\rme^{\thalf\rmi(\zeta-\zeta')^2/\tau}\;
{\sum_n}'\int_{\kappa_n(\nu)}^{\infty}d\kappa_y\; \rme^{\rmi\kappa_y(\eta-\eta')
-\rmi\varepsilon_n(\kappa_y)\tau}\nonumber\\
\fl\times \frac{D^2_{\varepsilon_n(\kappa_y)-1/2}\left[\sqrt{2}(\xi-\kappa_y)\right]}
{\int_0^{\infty}d\xi''\; D^2_{\varepsilon_n(\kappa_y)-1/2}
\left[\sqrt{2}(\xi''-\kappa_y)\right]}\; 
F\left(\sqrt{\tau[\nu-\varepsilon_n(\kappa_y)]},\frac{\zeta-\zeta'}{\sqrt{2\tau}}\right)
\label{8.4}
\end{eqnarray}
where $\kappa_n(\nu)$ is defined by the condition
$\varepsilon_n[\kappa_n(\nu)]=\nu$ with $n\leq \nu-\half$. In an analogous way,
the function $G$ for $x=x'$ is found as
\begin{eqnarray}
\fl G({\bf r},{\bf r}',t)=\frac{(1-\rmi)B^{3/2}}{4\pi^{3/2}\tau^{1/2}}\;
\rme^{\thalf\rmi(\zeta-\zeta')^2/\tau}\;
\sum_{n=0}^{\infty}
\int_{-\infty}^{\infty}d\kappa_y\; \rme^{\rmi\kappa_y(\eta-\eta')
-\rmi\varepsilon_n(\kappa_y)\tau}\nonumber\\
\times \frac{D^2_{\varepsilon_n(\kappa_y)-1/2}\left[\sqrt{2}(\xi-\kappa_y)\right]}
{\int_0^{\infty}d\xi''\; D^2_{\varepsilon_n(\kappa_y)-1/2}
\left[\sqrt{2}(\xi''-\kappa_y)\right]}
\label{8.5}
\end{eqnarray}
Here, the sum extends over all non-negative integer $n$ and the integral over all
$\kappa_y$, in contrast to the sum and integral in (\ref{8.4}).

For large values of $x$ the expressions (\ref{8.4}) and (\ref{8.5}) reduce to
their bulk forms (\ref{3.11}) and (\ref{3.12}). In fact, for
$\xi\rightarrow\infty$ only large values of $\kappa_y$ contribute to the
integrals. For these large $\kappa_y$ the function $\varepsilon_n(\kappa_y)$
reduces to $n+\half$. Furthermore, the parabolic cylinder function
$D_n(\sqrt{2}\, u)$ for integer $n$ can be rewritten in terms of a Hermite
polynomial as $2^{-n/2}\rme^{-u^2/2}H_n(u)$. Finally, the lower bounds of both
the integrals over $\xi''$ in the denominators in (\ref{8.4}), (\ref{8.5}) and
of the integral over $\kappa_y$ in (\ref{8.4}) can be extended to
$-\infty$. Evaluating the integrals with the use of (\ref{3.7}), one recovers
(\ref{3.11}) and (\ref{3.12}).

As a special case we consider the time correlation functions $G_{\mu}$ and $G$
for equal position vectors, as in section 3 and 7. One finds from (\ref{8.4})
and (\ref{8.5}):
\begin{eqnarray}
\fl G_{\mu}({\bf r},{\bf r},t)=\frac{B^{3/2}}{2\pi^{3/2}\tau^{1/2}}\;
{\sum_n}'\int_{\kappa_n(\nu)}^{\infty}d\kappa_y\; 
\rme^{-\rmi\varepsilon_n(\kappa_y)\tau}
\frac{D^2_{\varepsilon_n(\kappa_y)-1/2}\left[\sqrt{2}(\xi-\kappa_y)\right]}
{\int_0^{\infty}d\xi''\; D^2_{\varepsilon_n(\kappa_y)-1/2}
\left[\sqrt{2}(\xi''-\kappa_y)\right]}\nonumber\\
\times\left\{C\left(\sqrt{\tau[\nu-\varepsilon_n(\kappa_y)]}\right)-\rmi
S\left(\sqrt{\tau[\nu-\varepsilon_n(\kappa_y)]}\right)\right\}
\label{8.6}
\end{eqnarray}
and 
\begin{eqnarray}
\fl G({\bf r},{\bf r},t)=\frac{(1-\rmi)B^{3/2}}{4\pi^{3/2}\tau^{1/2}}\;
\sum_{n=0}^{\infty}
\int_{-\infty}^{\infty}d\kappa_y\; 
\rme^{-\rmi\varepsilon_n(\kappa_y)\tau}\nonumber\\
\times \frac{D^2_{\varepsilon_n(\kappa_y)-1/2}\left[\sqrt{2}(\xi-\kappa_y)\right]}
{\int_0^{\infty}d\xi''\; D^2_{\varepsilon_n(\kappa_y)-1/2}
\left[\sqrt{2}(\xi''-\kappa_y)\right]}
\label{8.7}
\end{eqnarray}
These expressions are the counterparts of (\ref{3.22}), (\ref{3.23}) for the
bulk, and of (\ref{7.4}), (\ref{7.5}) for the other geometry. For large $t$ the
Fresnel integrals in (\ref{8.6}) can be replaced by their asymptotic value
$\half$.

\begin{figure}
  \begin{center}
    \includegraphics[height=5.cm]{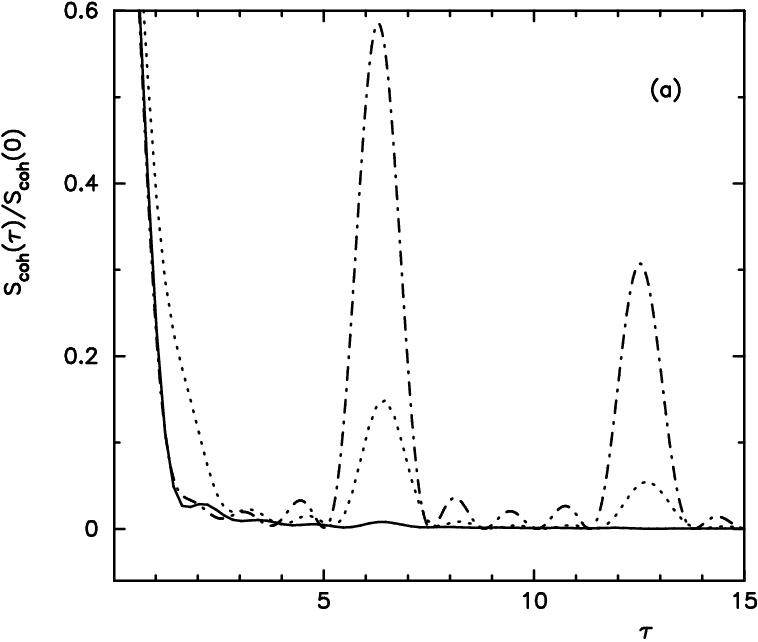}
    \hspace{0.5cm}
    \includegraphics[height=5.0cm]{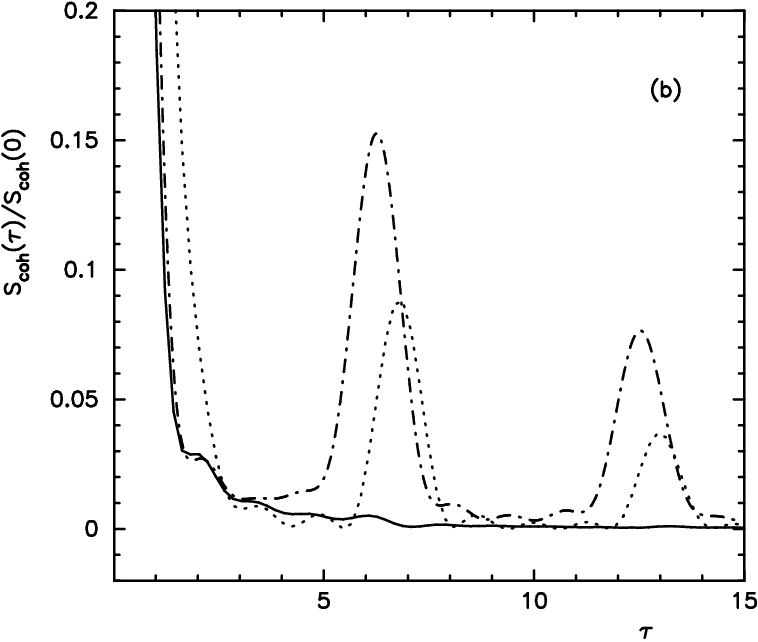}
\end{center}
\begin{center}
    \includegraphics[height=5.cm]{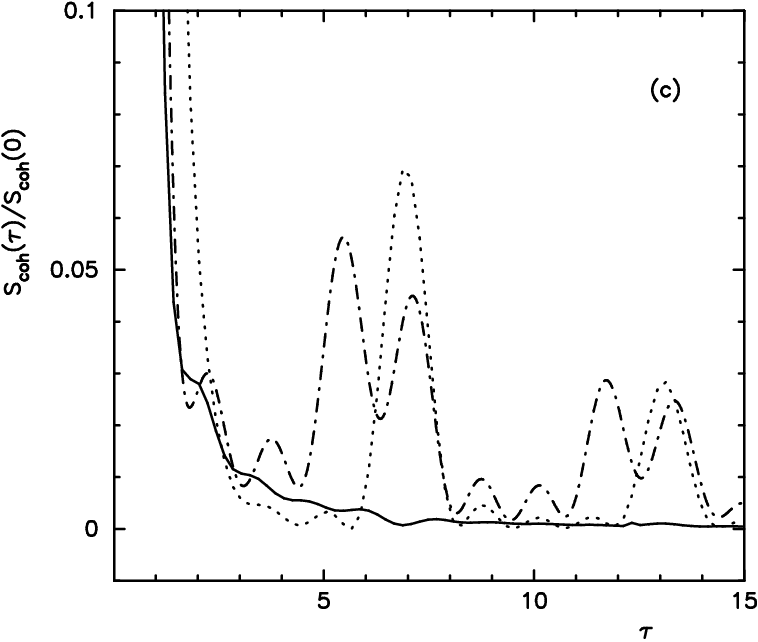}
    \hspace{0.5cm}
    \includegraphics[height=5.0cm]{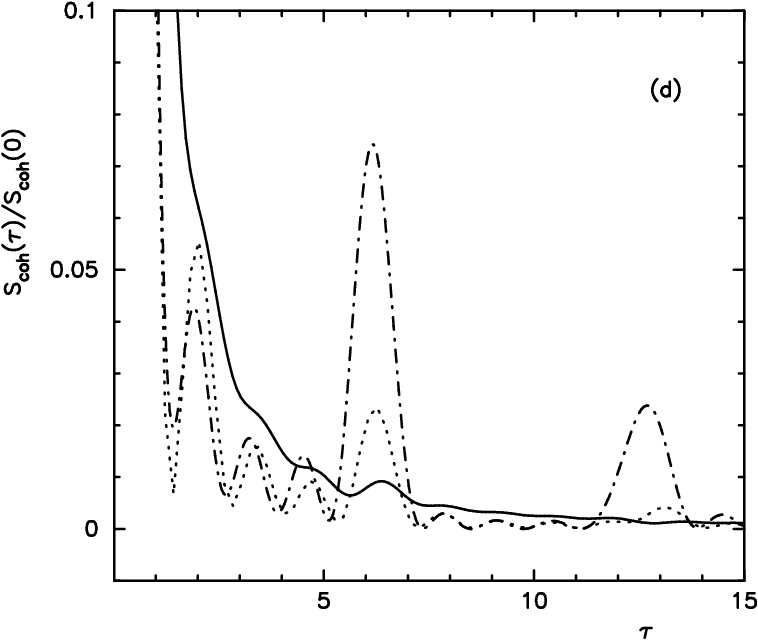}
  \caption{Coherent structure function as a function of time $\tau$, at
    $\nu=5$, for position difference $|\brho-\brho'|=1$ (a--c), and for
    coinciding positions (d). The first three correspond to 
    $\theta$ equal to $0$ (a), $\pi/4$ (b), $\pi/2$ (c), with
    $\theta=\arctan(|\eta-\eta'|/|\zeta-\zeta'|)$. The curves, which start at $1$
    for $\tau=0$, give the normalized structure function for distances $\xi$
    from the wall equal to $0.1$ (\full), $1.5$ (\dotted) and $\infty$
    (bulk, \chain).}
\label{fig5}
  \end{center}
\end{figure}

In figure \ref{fig5} numerical results for the normalized coherent structure
function, which follows by inserting (\ref{8.4}) or (\ref{8.6}) in (\ref{2.4}), 
are presented. Curves are drawn for several values of the distance $\xi$ from
the wall, and for various position differences, with a few different
orientations of the latter. The general picture of these figures is the same as
that obtained for the other geometry: near the wall the time decay of the
structure function is considerably faster than in the bulk. More details on
these figures will be given below, when discussing the time correlation
functions near the wall.

For $t=0$ the correlation function $G_{\mu}$ can be simplified by using the
asymptotic relation (\ref{3.15}). One gets from (\ref{8.4}):
\begin{eqnarray}
\fl G_{\mu}({\bf r},{\bf r}',t=0)=\frac{B^{3/2}}{2\pi^{2}|\zeta-\zeta'|}\;
{\sum_n}'\int_{\kappa_n(\nu)}^{\infty}d\kappa_y\;
\rme^{\rmi\kappa_y(\eta-\eta')}
\nonumber\\
\fl\times \frac{D^2_{\varepsilon_n(\kappa_y)-1/2}\left[\sqrt{2}(\xi-\kappa_y)\right]}
{\int_0^{\infty}d\xi''\; D^2_{\varepsilon_n(\kappa_y)-1/2}
\left[\sqrt{2}(\xi''-\kappa_y)\right]}\; 
\sin\left(\sqrt{2[\nu-\varepsilon_n(\kappa_y)]}\;|\zeta-\zeta'|\right)
\label{8.8}
\end{eqnarray}
which agrees with a result from \cite{KES:2001}, if spin degeneracy is taken
into account. As before, the second correlation function $G({\bf r},{\bf
r}',t)$ becomes a spatial delta function in the static limit.

As in the previous section, we want to study the behavior of the time
correlation functions in the vicinity of the wall. To that end we start from
(\ref{8.4}) and (\ref{8.5}), and use the identity \cite{KES:2001}
\begin{eqnarray}
\fl\frac{1}{\int_0^{\infty}d\xi''\; D^2_{\varepsilon_n(\kappa)-1/2}
\left[\sqrt{2}(\xi''-\kappa)\right]}=
-\frac{1}{2\pi}\; \Gamma^2[-\varepsilon_n(\kappa)+\half]\; 
D^2_{\varepsilon_n(\kappa)-1/2}\left(\sqrt{2}\,\kappa\right)\; 
\frac{d\varepsilon_n(\kappa)}{d\kappa}\nonumber\\
\mbox{}
\label{8.9}
\end{eqnarray}
Furthermore, we make a Taylor expansion of the parabolic cylinder function in
the numerators of (\ref{8.4}) and (\ref{8.5}). Subsequently, we employ the
Wronskian identity \cite{ERD:1953}
\begin{equation}
\fl \left. D_{\varepsilon_n(\kappa)-1/2}\left(\sqrt{2}\,\kappa\right)\; 
\frac{\partial}{\partial \kappa}
D_{\nu}\left(-\sqrt{2}\,\kappa\right)\right|_{\nu=\varepsilon_n(\kappa)-1/2}=
\frac{2\sqrt{\pi}}{\Gamma[-\varepsilon_n(\kappa)+\half]}
\label{8.10}
\end{equation}
which holds as a consequence of (\ref{8.2}). Upon changing the integration
variable from $\kappa_y$ to $\varepsilon$ in (\ref{8.4}), we finally obtain the
time correlation function for positions near the wall as
\begin{eqnarray}
\fl G_{\mu}({\bf r},{\bf r}',t)\approx\frac{\xi^2\; B^{3/2}}{2\pi^{3/2}\tau^{1/2}}\;
\rme^{\thalf\rmi(\zeta-\zeta')^2/\tau}\nonumber\\
\times{\sum_n}'\int_{n+\thalf}^{\nu}d\varepsilon \; \rme^{\rmi\kappa_n(\varepsilon)(\eta-\eta')
-\rmi\varepsilon\tau}\; 
F\left[\sqrt{\tau(\nu-\varepsilon)},\frac{\zeta-\zeta'}{\sqrt{2\tau}}\right]
\label{8.11}
\end{eqnarray}
Likewise, (\ref{8.5}) gives
\begin{equation}
\fl G({\bf r},{\bf r}',t)\approx\frac{(1-\rmi)\xi^2B^{3/2}}{2\pi^{3/2}\tau^{1/2}}\;
\rme^{\thalf\rmi(\zeta-\zeta')^2/\tau}\;
\sum_{n=0}^{\infty}
\int_{n+\thalf}^{\infty}d\varepsilon\; \rme^{\rmi\kappa_n(\varepsilon)(\eta-\eta')
-\rmi\varepsilon\tau}
\label{8.12}
\end{equation}
These expressions may be compared to (\ref{7.7}) and (\ref{7.8}) for the other
geometry. We will now consider the time correlation functions in the vicinity of
the wall for some particular orientations of the position difference, and for the
case of coinciding positions.

\subsection{Longitudinal position difference ($y=y',\; z\neq z'$)}
Putting $\eta=\eta'$ in the expression (\ref{8.11}) for $G_{\mu}$, we can
evaluate the integral by a partial integration. The result is
\begin{eqnarray}
\fl G_{\mu}({\bf r},{\bf r}',t)\approx 
-\frac{\rmi\xi^2\; B^{3/2}}{2\pi^{3/2}\tau^{3/2}}\;
\rme^{\thalf\rmi(\zeta-\zeta')^2/\tau}
{\sum_n}'\left\{\rme^{-\rmi(n+\thalf)\tau} \; 
F\left[\sqrt{\tau(\nu-n-\thalf)},\frac{\zeta-\zeta'}{\sqrt{2\tau}}\right]
\right.\nonumber\\
\left.-2\sqrt{\frac{\tau}{\pi}}\;
\rme^{-\rmi\nu\tau-\thalf\rmi(\zeta-\zeta')^2/\tau}\;
\frac{\sin\left[\sqrt{2(\nu-n-\thalf)}\;|\zeta-\zeta'|\right]}{|\zeta-\zeta'|}\;
\right\}
\label{8.13}
\end{eqnarray}
For large separations $|\zeta-\zeta'|$ we obtain with the use of (\ref{3.15}):
\begin{equation}
\fl G_{\mu}({\bf r},{\bf r}',t)\approx -\frac{\xi^2\; B^{3/2}}{\pi^2|\zeta-\zeta'|^2}\;
\rme^{-\rmi\nu\tau}{\sum_n}'\sqrt{2(\nu-n-\thalf)}\; 
\cos\left[\sqrt{2(\nu-n-\thalf)}\;|\zeta-\zeta'|\right]
\label{8.14}
\end{equation}
For $t=0$ the expressions (\ref{8.13}) and (\ref{8.14}) reduce to static results found
before \cite{KES:2001}. For large $\tau$ the second term between the curly
brackets in (\ref{8.13}) dominates, so that one gets:
\begin{equation}
\fl G_{\mu}({\bf r},{\bf r}',t)\approx 
\frac{\rmi\xi^2\; B^{3/2}}{\pi^2\tau}\;
\rme^{-\rmi\nu\tau}
{\sum_n}'\;\frac{\sin\left[\sqrt{2(\nu-n-\thalf)}\;|\zeta-\zeta'|\right]}{|\zeta-\zeta'|}
\label{8.15}
\end{equation}
Upon comparing this result with (\ref{7.11}), which holds for the other
geometry, one finds a similar behavior as a function of time: both have a decay
proportional to $\tau^{-1}$, which differs from the $\tau^{-1/2}$ decay of
$G_{\mu}$ for the bulk. These changes in the behavior of $G_{\mu}$ entail
modifications in the decay properties of the coherent structure function that
have been discussed in the previous section. The differences in the spatial
dependence between (\ref{7.11}) and (\ref{8.15}) are analogous to those found
in the bulk static correlation function (\ref{3.16}) for transverse and
longitudinal directions of the position difference, respectively.

Numerical results for the coherent structure function with a purely longitudinal
position difference are given in part (a) of figure \ref{fig5}. For a small distance
from the wall ($\xi=0.1$) the structure function decays rapidly. It does not
show the revivals around $\tau=2\pi m$, which are a prominent feature of the
bulk structure function, and which we have seen already in figure \ref{fig1}. For
intermediate distances ($\xi=1.5$ in the figure) these revivals have lost
part of their strength.

As to $G$, the integral in (\ref{8.12}) becomes trivial for $\eta=\eta'$, so
that one gets
\begin{equation}
 G({\bf r},{\bf r}',t)\approx
-\frac{(1-\rmi)\xi^2 B^{3/2}}{4\pi^{3/2}\tau^{3/2}\sin(\thalf\tau)}\;
\rme^{\thalf\rmi(\zeta-\zeta')^2/\tau}
\label{8.16}
\end{equation}
which may be compared to (\ref{7.8}) for the other geometry, and to (\ref{3.21})
for the bulk. As before, the time dependence shows the usual characteristic
recurring singularity and a faster decay than in the bulk.

\subsection{Transverse position difference ($z=z',\; y\neq y'$)}
In this case, the expression (\ref{8.11}) for $G_{\mu}$ becomes
\begin{eqnarray}
\fl G_{\mu}({\bf r},{\bf r}',t)\approx
\frac{\xi^2\; B^{3/2}}{\pi^{3/2}\tau^{1/2}}\;
{\sum_n}'\int_{n+\thalf}^{\nu}d\varepsilon \; 
\rme^{\rmi\kappa_n(\varepsilon)(\eta-\eta')-\rmi\varepsilon\tau}\nonumber\\
\times\left\{C\left[\sqrt{\tau(\nu-\varepsilon)}\right]-\rmi
S\left[\sqrt{\tau(\nu-\varepsilon)}\right]\right\}
\label{8.17}
\end{eqnarray}
In general, the integral cannot be simplified further. However, for large
$|\eta-\eta'|$ one may find its asymptotic behavior, as it is determined by the
upper boundary. One gets
\begin{equation}
\fl G_{\mu}({\bf r},{\bf r}',t)\approx
\frac{\xi^2\; B^{3/2}}{\sqrt{2}\pi^{3/2}}\; \rme^{-\rmi\nu\tau}\;
\frac{\rme^{\tthreequarter\pi\rmi\sgn{(\eta-\eta')}}}{|\eta-\eta'|^{3/2}}
{\sum_n}'\rme^{\rmi\kappa_n(\nu)(\eta-\eta')}
\left[-\frac{d\kappa_n(\nu)}{d\nu}\right]^{-3/2}
\label{8.18}
\end{equation}
which for $t=0$ reduces to an earlier result \cite{KES:2001}. As in (\ref{8.14})
the time dependence is given by a trivial phase factor. The space dependence is
quite different from that of (\ref{8.14}). 

For $G$ as given by (\ref{8.12}), the choice $\zeta=\zeta'$ produces only a
slight simplification. The integral in (\ref{8.12}) remains as it stands; it is
difficult to evaluate, even for large $|\eta-\eta'|$.

In part (c) of figure \ref{fig5} some numerical results for the coherent
structure function with a purely transverse position difference are
presented. As before, the structure function decays rapidly, if the distance
from the wall is small ($\xi=0.1$). The behavior is rather similar to that for
the other geometry (see part (a) of figure \ref{fig4}). At intermediate
distances ($\xi=1.5$) some revivals are present, as in the longitudinal case
(see part (a) of figure \ref{fig5}), albeit with a smaller amplitude (note the
difference in scale). At this distance from the wall the structure function is
quite different for the two geometries, as can be seen by comparing with the
corresponding curve in part (a) of figure \ref{fig4}. In the bulk the revivals
have a more complicated structure, as we have seen already in figure \ref{fig1}
and in figure \ref{fig4}. The intermediate case of a position difference that is
oriented at an angle $\pi/4$ with respect to the field leads to a behavior of
the coherent structure function that interpolates between the purely
longitudinal and the purely transverse cases, as can be seen in part (b) of
figure \ref{fig5}.

\subsection{Coinciding positions ($y=y',\; z=z'$)}
For coinciding positions we get from (\ref{8.13})
\begin{eqnarray}
\fl G_{\mu}({\bf r},{\bf r},t)\approx
-\frac{\rmi \xi^2 B^{3/2}}{\pi^{3/2} \tau^{3/2}} \;
{\sum_n}' \left\{
\rme^{-\rmi(n+\thalf)\tau}\;\left[ 
C\left(\sqrt{\tau(\nu-n-\thalf)}\right)\right.\right.\nonumber\\
\left.\left.-\rmi S\left(\sqrt{\tau(\nu-n-\thalf)}\right)\right]
-\sqrt{\frac{2}{\pi}}\;\rme^{-\rmi\nu\tau}\; \sqrt{\tau(\nu-n-\thalf)}\right\}
\label{8.19} 
\end{eqnarray}
Remarkably enough, this expression is identical to (\ref{7.9}), apart from the
interchange of $\xi$ and $\zeta$. Hence, near a wall the time correlation
function for equal positions has a universal form, independent of the orientation
of the wall with respect to the field.

The same conclusion is reached for $G$. Indeed, from (\ref{8.16}) we find for
$\zeta=\zeta'$:
\begin{equation}
G({\bf r},{\bf r},t)\approx
-\frac{(1-\rmi)\xi^2 B^{3/2}}{4\pi^{3/2}\tau^{3/2}\sin(\thalf\tau)}\;
\label{8.20}
\end{equation}
which has the same form as (\ref{7.10}).

For large $\tau$ the second term between the curly brackets in (\ref{8.19})
dominates. Introducing the bulk particle density, we may write
$G_{\mu}$ for coinciding positions near the wall in this asymptotic time regime
in a form analogous to (\ref{7.12}):
\begin{equation}
G_{\mu}({\bf r},{\bf r},t)\approx
\frac{2\rmi\xi^2}{\tau}\;\rme^{-\rmi\nu\tau}\; n_{\infty}
\label{8.21} 
\end{equation}
In accordance with the more general result discussed below (\ref{8.19}), the
long-time asymptotic behavior of the equal-position time correlation function near a
wall is independent of the orientation of the wall with respect to the
field. The asymptotic form has an even stronger universality property: it is
independent of the strength of the field as well. This follows by reintroducing 
the original (non-scaled) space and time variables, and the chemical potential
$\mu$. In terms of these variables (\ref{8.21}) reads:
\begin{equation}
G_{\mu}({\bf r},{\bf r},t)\approx \frac{2\rmi x^2}{t}\; \rme^{-\rmi\mu t}\; n_{\infty}
\label{8.22}
\end{equation}
The right-hand side is indeed independent of the magnetic field (apart from the
implicit dependence through the bulk particle density). It should be noted that
in the bulk the long-time tail in $G_{\mu}$ for equal positions does depend on
the magnetic field, as we have seen in (\ref{3.24}).

The universality property found for $G_{\mu}$ does not hold for $G$. In fact,
reintroducing the original variables in (\ref{8.20}) we get
\begin{equation}
G({\bf r},{\bf r},t)\approx
-\frac{(1-\rmi)x^2 B}{4\pi^{3/2}t^{3/2}\sin(\thalf Bt)}\;
\label{8.23}
\end{equation}
Clearly, this result depends on the magnetic field strength. It shows the
same characteristic recurrent pole structure as we have found in the bulk (see
(\ref{3.23})). As in the bulk, the poles at $t=2\pi n/B$ move when the field
strength changes. 

Numerical results for the coherent structure function with coinciding positions
are given in part (d) of figure \ref{fig5}. In the vicinity of the wall
($\xi=0.1$) the structure function is (almost) identical to that near a wall
perpendicular to the field (see the curve for $\zeta=0.1$ in part (b) of figure
\ref{fig4}), in accordance with the analytical results (\ref{7.9}) and
(\ref{8.19}). For intermediate distances from the wall the curves for the two
geometries are rather different, while they agree again in the bulk, of course.

\section{Concluding remarks}
In this paper we have studied time-dependent pair correlations in a quantum gas
of charged free fermions in a uniform external field. These correlations can be
described in terms of two correlation functions $G_{\mu,T}({\bf r},{\bf r}',t)$
and $G({\bf r},{\bf r}',t)$, which together determine both the coherent and the
incoherent part of the time-dependent structure function $S({\bf r},{\bf
r}',t)$, according to (\ref{2.3})--(\ref{2.6}). Our main interest has been
focused on the behavior of these correlation functions for the completely
degenerate quantum gas at temperature $T=0$.

Owing to the presence of the magnetic field, the particle motion is quantized in
terms of Landau levels with an associated cyclotron frequency. One of the aims
of the present paper has been to show how the correlation functions are
influenced by this cyclotron movement. The results (\ref{3.11}) and (\ref{3.12})
express the functions $G_{\mu,T=0}$ and $G$ as sums over Landau levels,
involving Laguerre polynomials for the spatial dependence and Fresnel integrals
for the time dependence. Their time behavior is sensitive to the orientation of
the position difference with respect to the magnetic field, as is demonstrated
in (\ref{3.17})--(\ref{3.21}). The time behavior of the ensuing coherent part of
the structure function, as shown in figure \ref{fig1}, is characterized by
recurrent `revivals', with a markedly increased strength of the correlation
effects at values of the time $t$ equal to a multiple of the inverse cyclotron
frequency. These revivals are particularly clear for position differences in the
longitudinal direction. The incoherent part of the structure factor is
influenced even more strongly: it is singular for the values of $t$ mentioned
above, as can be seen from (\ref{3.14}).

The time dependence of the correlation functions and the structure function can
be translated to a frequency dependence. Passing via the intermediate scattering
function $S({\bf k},t)$, for which we have derived the expressions (\ref{4.8})
and (\ref{4.18}), we arrive at the dynamical structure factor $S({\bf
k},\omega)$, as given by (\ref{5.1}) and (\ref{5.3}). The curves in figure
\ref{fig2} and \ref{fig3} show how both the coherent and the incoherent part of
$S({\bf k},\omega)$ get a spike structure for purely transverse orientations of
the wavevector. The gradual change of these curves as the wavevector gets a
longitudinal component can be inferred from the figures. For a purely
longitudinal wavevector a step structure is found, in accordance with
(\ref{5.5}).

The above results pertain to the bulk of magnetized quantum gases. As is well
known, these systems show additional interesting physical phenomena near the
walls of the containers to which they are confined. A second goal of the present
paper has been to investigate how the time-dependent correlations change in the
vicinity of a confining wall. In general, these edge effects depend on the
orientation of the wall with respect to the field. They have been determined for
two specific orientations, namely for a wall perpendicular to the field and for
a wall parallel to the field. In the former case the cylinder symmetry of the
field is not changed by the wall. As a consequence, the time correlation
functions are easily obtained from those of the bulk, with results given in
(\ref{7.2})--(\ref{7.3}). The curves of figure \ref{fig4} show that the main
influence of the wall is a faster time decay of the correlations. For a wall
parallel to the field it is a lot more cumbersome to determine the correlation
functions, as the symmetry of the problem is now lost completely. The general
results for the correlation functions are given in (\ref{8.4})--(\ref{8.5}). The
curves for the coherent structure function, which are presented in figure
\ref{fig5}, show that once again the correlations die out faster than in the
bulk. In particular, the characteristic revivals at multiples of the inverse
cyclotron frequency, which are a conspicuous feature of the bulk coherent
structure function for longitudinal position differences, are almost absent for
positions near the wall. It appears that the wall perturbs the regular patterns
of the bulk motion so strongly that the recurrences associated to the cyclotron
frequency do no longer influence the correlation functions.

\appendix
\section*{Appendix}
\setcounter{section}{1}
In this appendix we consider several integral relations that are employed in the
main text. The first identity, which has been used in section 4, is
\begin{eqnarray}
\fl \int_{-\infty}^{\infty}dz \; \rme^{-\rmi kz}\;
F^{\ast}(a,zc)\; F(b,zc)\nonumber\\
\fl = -\frac{2\rmi}{k}\; \rme^{\tfourth \rmi k^2/c^2}\;
\left\{\rme^{\rmi kb/c}\; \theta\left[4a^2c^2-(k+2bc)^2\right]-
\rme^{-\rmi kb/c}\; \theta\left[4a^2c^2-(k-2bc)^2\right]\right\}\nonumber\\
\fl +\frac{2\rmi}{k}\; \rme^{-\tfourth \rmi k^2/c^2}\;
\left\{\rme^{-\rmi ka/c}\; \theta\left[4b^2c^2-(k+2ac)^2\right]-
\rme^{\rmi ka/c}\; \theta\left[4b^2c^2-(k-2ac)^2\right]\right\}
\label{A.1}
\end{eqnarray}
with the function $F$ as defined in (\ref{3.10}) and with positive $a$, $b$, $c$
and real $k\neq 0$. To prove it, one starts by using a partial integration to
rewrite the left-hand side as
\begin{eqnarray}
\fl -\frac{\rmi\,\sqrt{2}\, c}{\sqrt{\pi}\, k}\int_{-\infty}^{\infty}dz\; 
\rme^{-\rmi kz}\left\{
\left[\rme^{\rmi(a+zc)^2}-\rme^{\rmi(a-zc)^2}\right]\; F(b,zc)\right.\nonumber\\
+\left.\left[\rme^{-\rmi(b+zc)^2}-\rme^{-\rmi(b-zc)^2}\right]\; F^{\ast}(a,zc)\right\}
\label{A.2}
\end{eqnarray}
To determine the integral at the right-hand side, we have to evaluate integrals
of the form:
\begin{equation}
G(k,a,b)=\int_{-\infty}^{\infty}dz\; \rme^{-\rmi kz+\rmi b^2 z^2}\; F(a,zb)
\label{A.3}
\end{equation}
for positive $a$, $b$ and real $k$. We differentiate this function with respect
to $k$. By adding a suitable multiple of $G$ itself we arrive at:
\begin{equation}
-2b^2\; \frac{\partial G(k,a,b)}{\partial k}-\rmi k\; G(k,a,b)=
\int_{-\infty}^{\infty}d\left(\rme^{-\rmi kz +\rmi b^2 z^2}\right)\; F(a,zb)
\label{A.4}
\end{equation}
Integrating by parts and evaluating the ensuing integral the right-hand side
becomes a combination of delta functions:
\begin{equation} 
- 2^{3/2}\; \sqrt{\pi}\; b\; \rme^{-\rmi a^2}\; \left[ 
\delta(k+2ab)-\delta(k-2ab)\right]
\label{A.5}
\end{equation}
By combining (\ref{A.4}) and (\ref{A.5}) we arrive at a differential equation,
which can easily be solved as:
\begin{equation}
G(k,a,b)=\frac{\sqrt{2\pi}}{b}\; \rme^{-\tfourth\rmi k^2/b^2}\; 
\theta\left(4 a^2 b^2-k^2\right)+\bar{G}(a,b)\; \rme^{-\tfourth\rmi k^2/b^2}
\label{A.6}
\end{equation}
with an as yet unknown function $\bar{G}(a,b)$. To determine it we put $k=0$,
and differentiate both sides with respect to $a$. Using the definition
(\ref{A.3}) we find, by differentiating $F(a,zb)$ and performing the resulting
integral, that $G(0,a,b)$ is independent of $a$ for all positive $a$ and $b$. As
a consequence $\bar{G}(a,b)$ is independent of $a$ as well. Finally, by
considering the limit $a\rightarrow \infty$ (for $k=0$ as before) in (\ref{A.3})
and employing (\ref{3.15}), we prove
$\lim_{a\rightarrow\infty}G(0,a,b)=\sqrt{2\pi}/b$, and hence
$\lim_{a\rightarrow\infty}\bar{G}(a,b)=0$. From these results we conclude that
the term with $\bar{G}$ in (\ref{A.6}) can be omitted, so that $G(a,b)$ is
determined. After these preparations it is an easy matter to demonstrate the
validity of (\ref{A.1}).

A second identity (used in section 6) is 
\begin{eqnarray} 
\fl \int_0^a dv\; \rme^{-\rmi vt}\; J_0\left(\sqrt{v}\, b\right)\;
F\left(\sqrt{(a-v)t}\, ,\frac{c}{2\sqrt{t}}\right)
= -\frac{\rmi}{t}\; \rme^{\tfourth\rmi b^2/t}\;
F\left(\sqrt{at},\frac{\sqrt{b^2+c^2}}{2\sqrt{t}}\right) \nonumber\\
+\frac{2^{3/2}\rmi}{\sqrt{\pi t}\, \sqrt{b^2+c^2}}\; 
\rme^{-\rmi at-\tfourth\rmi c^2/t}\; \sin\left[\sqrt{a(b^2+c^2)}\right] 
\label{A.7} 
\end{eqnarray} 
for non-negative $b$, $c$ (with $\sqrt{b^2+c^2}>0$) and positive $a$, $t$. Before
starting with the proof of this identity, we remark that its dependence on $a$
is trivial, as follows from a rescaling of the variables. Hence, we will put
$a=1$ in the following. The function $F$ in the integrand at the left-hand side
of (\ref{A.7}) may be replaced by the integral representation
\begin{equation}
F\left(\sqrt{(1-v)t}\, ,\frac{c}{2\sqrt{t}}\right)=\sqrt{\frac{2t}{\pi}}\;
\rme^{-\tfourth\rmi c^2/t}\;\int_{-\sqrt{1-v}}^{\sqrt{1-v}}dk\; 
\rme^{-\rmi ck-\rmi t k^2}
\label{A.8}
\end{equation}
as may be established from the definition of the Fresnel integrals (or by
comparing (\ref{3.6}) and (\ref{3.9})). Inserting the integral representation in
(\ref{A.7}), we arrive at a double integral over $v$ and $k$. Introducing new
integration variables $\rho$ and $\theta$ by writing $v=\rho^2\sin^2\theta$ and
$k=\rho\cos\theta$, we rewrite this double integral as
\begin{equation}
2\int_0^1 d\rho\; \rho^2\; \rme^{-\rmi t\rho^2}\;
\int_0^{\pi}d\theta\; \sin\theta\; J_0(b\rho\sin\theta)\; 
\rme^{-\rmi c\rho\cos\theta}
\label{A.9}
\end{equation}
The integral over $\theta$ turns out to depend on $b$ and $c$ only through the
combination $\sqrt{b^2+c^2}$. This can be seen by using the standard
representation for the Bessel function and writing the $\theta$-integral as 
\begin{equation}
\frac{1}{2\pi}\; \int_0^{\pi}d\theta\;\int_0^{2\pi}d\varphi\; \sin\theta\; 
\rme^{-\rmi b\rho\, \sin\theta\, \cos\varphi -\rmi c\rho\, \cos\theta}
\label{A.10}
\end{equation}
The argument of the exponential function is proportional to the scalar product
of a unit vector in the direction of $(\theta,\varphi)$ and a constant vector
with cartesian components $(b,0,c)$. Since the integration extends over all
directions of the unit vector, the integral (\ref{A.10}), and hence (\ref{A.9})
as well, indeed depends on $b$ and $c$ through the length $\sqrt{b^2+c^2}$
only. Once this is known, we may evaluate (\ref{A.9}) by choosing $b=0$ and
replacing $c$ by $\sqrt{b^2+c^2}$ in the answer. Since the Bessel function
equals 1 for $b=0$, the integral (\ref{A.9}) can easily be evaluated in terms of
Fresnel integrals, with the result:
\begin{equation}
\frac{2\rmi}{ct}\; \rme^{-\rmi t}\; \sin c -\frac{\rmi\sqrt{\pi}}{\sqrt{2}\,
t^{3/2}}\;\rme^{\tfourth \rmi c^2/t}\; F\left(\sqrt{t}\,
,\frac{c}{2\sqrt{t}}\right)
\label{A.11}
\end{equation}
Generalizing this answer to the case $b\neq 0$ in the way indicated, and
substituting it in the left-hand side of (\ref{A.7}), we arrive at the
desired result.

\end{document}